\documentstyle[12pt]{report}
\def\bcn{\begin{center}}
\def\ecn{\end{center}}
\def\citenum#1{{\def\@cite##1##2{##1}\cite{#1}}}
\newcommand{\beqn}{\begin{eqnarray}}
\newcommand{\eeqn}{\end{eqnarray}}
\def\spose#1{\hbox to 0pt{#1\hss}}
\def\lsim{\mathrel{\spose{\lower 3pt\hbox{$\mathchar"218$}}
     \raise 2.0pt\hbox{$\mathchar"13C$}}}
\def\gsim{\mathrel{\spose{\lower 3pt\hbox{$\mathchar"218$}}
     \raise 2.0pt\hbox{$\mathchar"13E$}}}
\def\simpropto{\mathrel{\spose{\lower 3pt\hbox{$\mathchar"218$}}
     \raise 2.0pt\hbox{$\propto$}}}
\def\beq{\begin{equation}}
\def\eeq{\end{equation}}
\def\barr{\begin{array}}
\def\earr{\end{array}}
\def\ibid{{\sl ibid.}}

\def\tr{{\rm tr\,}}
\def\hc{\rm H.c.}
\def\diag{\rm diag}
\def\calo{{\cal O}}
\def\calu{{\cal U}}
\def\calv{{\cal V}}
\def\calq{{\cal Q}}
\def\calm{{\cal M}}
\def\cals{{\cal S}}
\def\sinb{\sin \beta}
\def\cosb{\cos \beta}
\def\tanb{\tan \beta}
\def\barv{\overline{v}}
\def\barm{\overline{m}}
\def\psibar{\overline{\psi}}
\def\vev#1{{\langle#1\rangle}}
\def\etal{ {\it et al.}}
\def\ie{ {\it i.e.} }
\def\eg{ {\it e.g.} }
\def\AJ#1#2#3{{\sl Astr. J.} {\bf #1}, #2 (#3)}
\def\ZPC#1#2#3{{\sl Z.~Phys.} {\bf C#1}, #2 (#3)}

\def\PRL#1#2#3{{\sl Phys. Rev. Lett.} {\bf #1}, #2 (#3)}
\def\PRD#1#2#3{{\sl Phys. Rev.} {\bf D#1}, #2 (#3)}
\def\PLB#1#2#3{{\sl Phys. Lett.} {\bf B#1}, #2 (#3)}
\def\PREP#1#2#3{{\sl Phys. Rep.} {\bf #1}, #2 (#3)}
\def\NPB#1#2#3{{\sl Nucl. Phys.} {\bf B#1}, #2 (#3)}

\def\gev{{\rm GeV }}
\def\Mev{{\rm MeV }}
\def\kev{{\rm keV }}
\def\ev{{\rm eV }}

\def\mgut{M_{\rm GUT}}

\def\mz{m_{\rm z}}
\def\mw{m_{\rm w}}
\def\xiw{\xi_{\rm w}}
\def\xiz{\xi_{\rm z}}

\def\abs#1{\left|#1\right|}

\def\fivebar{\overline{\bf 5}}

\def\drbar{\overline{\rm DR}}
\def\tanb{\tan\beta}

\def\half{{1\over 2}}

\def\ifmath#1{\relax\ifmmode #1\else $#1$\fi}
\def\ZPC#1#2#3{{\sl Z.~Phys.} {\bf C#1}, #2 (#3)}

\def\PRL#1#2#3{{\sl Phys. Rev. Lett.} {\bf #1}, #2 (#3)}
\def\PRD#1#2#3{{\sl Phys. Rev.} {\bf D#1}, #2 (#3)}
\def\PLB#1#2#3{{\sl Phys. Lett.} {\bf B#1}, #2 (#3)}
\def\PREP#1#2#3{{\sl Phys. Rep.} {\bf #1}, #2 (#3)}
\def\NPB#1#2#3{{\sl Nucl. Phys.} {\bf B#1}, #2 (#3)}

\def\tanw{\tan\theta_{\rm w}}
\def\sw{s_{\rm w}}
\def\cw{c_{\rm w}}
\def\sb{s_{\beta}}
\def\cb{c_{\beta}}

\def\gev{{\rm GeV }}

\def\mgut{M_{\rm GUT}}

\def\abs#1{\left|#1\right|}
\def\tanb{\tan\beta}

\def\d{{\rm d}}

\def\mpl{M_{\rm P}}

\def\half{\ifmath{{\textstyle{1 \over 2}}}}

\begin{document}
\thispagestyle{empty}
\setcounter{page}{0}

\begin{flushright}
MPI-PhT/95-59\\
NSF-ITP-95-148\\

November 11, 1995\\
revised: July 8, 1996
\end{flushright}
\bigskip
\bigskip
\bigskip
\bigskip
\bigskip

\begin{center}
{\Large\bf Neutrino Masses and Mixing Angles in
SUSY-GUT Theories with explicit R-Parity
Breaking
}\\
\vspace{2em}
\large
Ralf Hempfling

{\it Max-Planck-Institut f\"ur Physik,
Werner-Heisenberg-Institut,}

{\it D--80805 M\"unchen, Germany}

\vspace{1.5ex}
{\it Email:} {\tt hempf@mppmu.mpg.de}
\end{center}
\bigskip
\bigskip
\bigskip

\begin{abstract}
\noindent
In minimal SUSY GUT models the $R$-parity
breaking terms are severely constrained by
SU(5) gauge invariance. We consider the
particular case where
the explicit $R$-parity breaking occurs only
via dimension 2 terms of the superpotential.
This model possesses only three R-parity breaking parameters.
We have studied the predictions of
this model for the neutrino masses and mixing angles
at the one-loop level
within the framework of a radiatively broken
unified supergravity model.
We find that this model naturally yields masses and mixing angles
that can explain the solar and atmospheric neutrino
problems. In addition, there are regions in parameter space
where the solution to the solar neutrino puzzle
is compatible with either the LSND result or
the existence of significant hot dark matter neutrinos.

\end{abstract}
\newpage

\chapter{Introduction}

The standard model of elementary particle physics (SM)
is in very good agreement with all presently available data.
Nonetheless, it suffers from various theoretical shortcomings,
the most severe of which is the hierarchy problem\cite{hierarchy}.

In supersymmetric models the cancellation of quadratic divergences
is guaranteed and thus any mass scale is stable under radiative
corrections. Supersymmetry (SUSY) implies that any fermion (boson)
is accompanied by a bosonic (fermionic) superpartner
with the same mass and
transformation properties under the gauge symmetry\cite{susyreview}.
The most economical candidate for a realistic model
is the minimal supersymmetric extension of the SM (MSSM).
In addition to the superpartners for all SM particles
it contains two Higgs bosons (required to give mass to up and down-type 
fermions and to cancel the triangle anomalies arising form 
the fermionic partners of the Higgs bosons) but no other particle.

The most general superpotential
invariant under the gauge symmetry
can be written as
\beqn
W &=& \half y^L_{I J k} \hat L_I \hat L_J \hat E^c_k
   +      y^D_{I j k} \hat L_I \hat Q_j \hat D^c_k
   -      y^U_{  j k} \hat H   \hat Q_j \hat U^c_k
   -      \mu_I   \hat L_I \hat H
   + \half \bar y^D_{i j k}\hat  D^c_i\hat  D^c_j\hat  U^c_k\,,
\label{defw}
\eeqn
where the supermultiplets are denoted by a hat.
The left-handed lepton supermultiplets
are denoted by $\hat L_i$ ($i = 1,2,3$)
and the Higgs superfield coupling to the down-type quarks
is denoted by $\hat L_0$.
Throughout this paper, 
we use the notation $i,j,k = 1,2,3$ and $I,J,K = 0,1,2,3$
and we sum over twice occurring indices.
Note that $L_I L_J \equiv \epsilon_{a b} L_I^a L_J^b = - L_J L_I$
($a,b = 1,2,3$ are the SU(2)$_L$ indices)
and thus $y_{I J k}^L = - y_{J I k}^L$,

Let us first determine the meaning of the various terms of
eq.~\ref{defw}. Here, $y_{0jk}^L$, $y_{0jk}^D$ and $y_{jk}^U$
denote the lepton, down-type and up-type Yukawa couplings, respectively,
and $\mu_0$ is the Higgs mass parameter.
However, in contrast to the SM the MSSM allows for 
baryon [lepton] number violating interactions
$\bar y_{ijk}^D$ [$y_{ijk}^L$ and $\mu_i$].

These couplings are constrained from above by experiment.
The most model independent constraints can be obtained from 
collider experiments\cite{collider-c} or neutrino
physics [\citenum{neutrino-c1},\citenum{neutrino-c2}].
It turns out that the individual
lepton and baryon number
violating couplings only have to be smaller
than $O(10^{-3}\sim {\rm few}\times 10^{-1})$.
Thus, the $R$-parity violating couplings need not
 be more suppressed than the
lepton and baryon number preserving Yukawa couplings.
The exception is the constraint on the coupling
\beqn
\bar y_{121}^D\lsim 10^{-7}
\label{y121}
\eeqn
from heavy nuclei decay\cite{sher}, but even this constraint becomes
somewhat less
impressive if compared to $y^D_{011} \simeq 3\times 10^{-5}$.
Somewhat stronger constraints can be derived
from cosmology\cite{cosmology-c}.


Thus, it may be premature to conclude from our negative
experimental search that $R$-parity is a good
approximation or even exact symmetry of nature.

However, the experimental exclusion area can be strongly enhanced by
imposing theoretical constraints.
In the minimal SU(5) SUSY-GUT model, the right-handed leptons,
the right-handed up-type quarks and the left-handed quarks are embedded
in a ten dimensional representation,
${\bf 10}_i = E^c_i\oplus U^c_i \oplus Q_i$.
The right-handed down-type 
quarks and the left-handed leptons are embedded
in a five dimensional representation,
${\bf \overline{5}}_i = D_i\oplus L_i$.
The two Higgs doublets are embedded together
with two proton decay mediating
colored triplets, $T$ and $D_0$, in five dimensional representations,
${\bf \overline{5}}_0 = D_0\oplus L_0$ and
${\bf           5 }   = T\oplus H$.
In this model, both the lepton and the baryon number violating 
interactions arise from the term
\beqn
W_{\rm GUT} = \half y_{i j k} {\bf \overline{5}}_i 
{\bf \overline{5}}_j {\bf 10}_k
\,,\label{wsu5}
\eeqn
where the boundary conditions at $\mgut$ are given by
$y^L_{i j k} = y^D_{i k j} = \bar y_{i j k}^D = y_{i j k}$.\footnote{It
is clear that the predictions for the
down-type quark masses of the first two generation
in the minimal model are off by factors of $O(3)$.
Thus, any more realistic
model has to be more complicated. 
While it is easy to reconcile the mass predictions with experiment,
e.g. by introducing higher dimensional
representations or higher dimensional operators,
it would be very hard to explain why
either the lepton number violating Yukawa couplings or
the baryon number violating couplings are suppressed by many
orders of magnitude.}
Thus, in general the baryon and lepton number
violating couplings are correlated in SUSY GUT models.
This leads to very strong constraints
on any $y_{i j k}$ from proton
stability which are much stronger then 
the constraint in eq.~\ref{y121}\cite{smirnov}.
(Note, that it does not constraint the coefficients
of dimension 2 terms, $\mu_i$.)

These much stronger constraints in the framework of SUSY GUT models
are the reason why in the MSSM any lepton and baryon number violating
interaction is eliminated by imposing
a discrete, multiplicative symmetry called
$R$-parity\cite{r-parity}
\beqn
R_p = (-1)^{2S+3B+L}\,,
\eeqn
where $S$, $B$ and $L$ are the spin, baryon and lepton numbers,
respectively.
Aside from the long proton life-time,
$R$-parity conserving models have the very attractive feature
that the lightest supersymmetric particle (LSP)
is stable and a good cold dark matter candidate\cite{cdm}.
However, while the existence of a dark matter candidate
is a very desirable prediction, it does not prove
$R$-parity conservation and 
one should keep an open mind for more general models.

In this paper, we will investigate
the scenario where $R$-parity is broken explicitly via
$\mu_i \neq 0$\cite{suzuki}. 
In particular, we compute the predictions for the neutrino
masses and mixing angles
in the frame-work of a SUSY-GUT model with
radiative electroweak symmetry breaking.

Our paper is organized as follows: in section~2 we present the
model at tree-level and our notation and conventions. In section~3 
we present our numerical analysis of the one-loop radiatively
corrected neutrino/neutralino mass and mixing matrix. Our conclusions
are presented in section~4 and our results
for the renormalization group equations (RGEs), the expressions
for masses and vertices, and the formulae for our one-loop results
are relegated to four appendices.

\chapter{The Model at Tree-Level}

In the introduction we have argued that explicit 
$R$-parity breaking Yukawa
couplings are strongly constrained by the long proton life-time
if the model originates from a SUSY-GUT model.
One alternative is to break $R$-parity spontaneously
 at the electro-weak scale.
This possibility has been studied 
extensively [\citenum{neutrino-c2},\cite{rspontaneous1},\cite{rspontaneous2}].
The main feature of these models is the existence of a
Goldstone-boson, the Majoron, associated with 
the spontaneous breaking of
a continuous symmetry, Lepton number conservation.
The simplest model where $R$-parity is broken spontaneously by
a sneutrino VEV\cite{rspontaneous1} is phenomenologically ruled out. 
This leads to the introduction of several SM singlets
and new Yukawa-type couplings\cite{rspontaneous2}.
Another possibility is the explicit $R$-parity breaking via
soft terms of the superpotential only.
These terms are unconstrained
by the proton life-time
and can arise naturally in  SUSY-GUTs.

Let us consider, for example, a SUSY GUT model based on SO(10).
Here, the Higgs fields are embedded in a ${\bf 10} = 
{\bf 5} \oplus \fivebar_0$
while the down-type fermions are embedded together with 
all other fermions of
one generation including the right-handed neutrino in a 
${\bf 16} = {\bf 10} \oplus \fivebar \oplus {\bf 1}$.
The only renormalizable Yukawa couplings in SO(10) can be written as
$W_{GUT} = y_{i j} {\bf 10}\otimes{\bf 16}_i\otimes {\bf 16}_j$.
This automatically implies the absence of the couplings
$y_{i j k}^L$, $y_{i j k}^D$ and $\bar y_{i j k}^D$.
The only non-zero components are from now on
abbreviated by
$y_{i j}^L \equiv y_{0 i j}^L$ and
$y_{i j}^D \equiv y_{0 i j}^D$.

Furthermore, the Higgs/slepton mixing terms vanish (\ie\ $\mu_i = 0$).
However, in general this term can be generated dynamically via
$W_{mix} = y_{i} {\bf 10}\otimes{\bf 16}_i \otimes{\bf 16}_H$
if SO(10) is broken spontaneously
to SU(5) by the right-handed neutrino-like component
of the 16 dimensional Higgs field $\vev{{\bf 16}_H} \neq 0$.
Why the resulting mass terms $\mu_i \equiv \lambda_i\vev{{\bf 16}_H}$
and $\mu_0$ are of the order of $\mz$ rather
than the scale of gauge unification, $\mgut \simeq 10^{16}~\gev$
is not clear but possible mechanisms are
known\cite{mu-problem}.  It is not implausible that the couplings
$y_i$ arise from non-renormalizable terms and are $O(m_{3/2}/\mpl)$
(here, $m_{3/2}\simeq$ few$\times100~\gev$ is the gravitino mass
and $\mpl \simeq 10^{19}~\gev$ is the Planck mass).
Whatever the solution for the $\mu$ problem, it is expected to
affect the $\mu_i$ in an analogous fashion.
Thus we can expect naturally that $\mu_i = O(\mu_0)$\footnote{
the only plausible alternative would be $\mu_i = \mgut$
which is phenomenologically ruled out.}
with a possible suppression factor of 
$\vev{{\bf 16}_H}/\mpl\simeq 10^{-3}$.
In this model, the analogous baryon number
 violating Higgs triplet/down-type
quark mixing is suppressed automatically if
 $D_0$ (together with $T$) acquires a mass
$O(\mgut)$. Such a large mass for the Higgs triplets is 
required in any SUSY-GUT model with or without $R$-parity.
It is a very severe problem\cite{doublettriplet}
and has no completely satisfying solution to date.
However, once this problem, common to all SUSY-GUT models,
is solved there are no additional baryon number violating
terms in the low energy theory even if we include non-zero values
of $\mu_i$.

On the other hand, lepton number violation via Higgs/slepton mixing
is potentially unsuppressed. This gives rise to 
neutrino masses and mixing angles and may provide an
economical explanation for the solar neutrino
deficiency\cite{solarn} and to solve the atmospheric
neutrino puzzle\cite{atmosphericn}.
These possibilities are well within present constraints
and deserve to be studied.
Thus, we will focus our attention to the scenario where
$y_{i j k} = 0$ but $\mu_i \neq 0$.

\section{The SUSY-GUT framework}

In any realistic model
SUSY has to be broken at the electro-weak scale
given by the mass of the $Z$ boson, $\mz = 91.187$\cite{pdg}
in order to explain why no superpartner has
been detected to date.
Here, this is done 
by including explicit soft SUSY breaking terms in the potential
\beqn
V_{\rm soft} &=& \left[A^L_{i j} \widetilde L_0\widetilde L_i
         \widetilde E^c_j
   +      A^D_{i j}\widetilde L_0\widetilde Q_i\widetilde D^c_j
   -      A^U_{i j}H \widetilde  Q_i\widetilde U^c_j
   -      B^\mu_I   \widetilde L_I H + {\rm H.c.}\right]\nonumber\\
  &&+\sum_{m,n} M_{\phi_{m n}}^2\phi_m^{\dagger} \phi_n
    +\sum_a M_a   \lambda_a\lambda_a + \hc\,,
\label{defvsoft}
\eeqn
where $\phi_n = \widetilde L_I, \widetilde E_i, \widetilde Q_i,
\widetilde U_i, \widetilde D_i, H$ stands for all the sfermion fields
and the Higgs doublets.
The gauginos are denoted by $\lambda_a$
[the indices $a,b = 1,2,3$ refer to the U(1), SU(2)$_L$ and SU(3)$_c$
gauge symmetries, respectively; the gauge indices are suppressed].

The large number of free soft SUSY breaking parameters can
be reduced significantly by making certain assumptions 
about their origin.
In minimal $N= 1$ supergravity (SUGRA) models the breaking 
of SUSY is thought
to occur in a ``hidden'' sector \ie\ a sector of the theory that couples
to the standard model particles only via gravity.
The soft SUSY breaking terms arising from gravitational coupling to the
``hidden sector'' are assumed to be universal at the Planck scale
$\mpl \simeq 10^{19}~\gev$.
As a result, there are only four independent
parameters.
These are the $A_0$ ($B_0$) parameter multiplying the tri-linear (bilinear)
part of $W$, the universal scalar mass $m_0$ and the 
universal gaugino mass
parameter $m_{1/2}$, \ie\
\beqn
A^L_{I J k} &=& A_0 y^L_{I J k}\,, \qquad
A^D_{I j k}  =  A_0 y^D_{I j k}\,, \qquad
A^U_{i j}    =  A_0 y^U_{i j}\,,\nonumber\\
B^\mu_I &\equiv& B_{I J} \mu_J = B_0 \mu_I\,,\nonumber\\
M^2_{\phi_{m n}} &=& m_0^2 \delta_{m n}\,,\nonumber\\
M_a     &=& m_{1/2}\,.
\label{softbc}
\eeqn
By assuming universal soft SUSY breaking parameters at
$\mpl$ we have drastically reduced the number of free
SUSY breaking parameters of the theory. 
This universality of the soft SUSY breaking parameters is broken at 
the electro-weak scale due to a renormalization group
evolution of the individual parameters\footnote{%
Strictly speaking, this scenario assumes the absence of any
intermediate scale between $\mpl$
and $\mz$ since significant non-universal terms can arise already
in the relatively small range between $\mpl$ and $\mgut$
due to large group coefficients of a unified group\cite{nir+pomerol}.
However, these effects are strongly model dependent and shall 
be neglected
here.}
(the $\beta$ functions
for the general model with $R$-parity breaking
are listed in appendix~A).  It is one of the
great successes of this scenario that the low energy value of
$m_H^2$ obtained from renormalization group evolution 
is indeed negative,
giving rise to spontaneous electroweak symmetry breaking
while all other scalar mass parameters remain positive over a large
region of parameter space\cite{radssb}.

\section{Minimizing the Higgs Potential}

Before we can investigate any other sector of the model we 
first have to minimize the Higgs potential.
The fundamental difference of our model to the 
MSSM  with conserved $R$-parity
is the fact that the Higgs fields mix with
slepton fields. Thus, we effectively have to minimize a
five Higgs doublet model.
In particular, all the mass eigenstates and the vacuum expectation
values will be a linear combination of Higgs fields and
slepton fields. 
The tree-level potential can be written as
\beqn
V &=& (\mu^2 + m_H^2) H^\dagger H
   +(\mu_I \mu_J + m_{L_{I J}}^2) \widetilde L_I^\dagger \widetilde L_J
   -B_I^\mu \left(\widetilde L_I H+\hc\right)\nonumber\\
   && +{g^2+g^{\prime 2}\over 8}\left(H^\dagger H 
          - \widetilde L^\dagger_I \widetilde L_I\right)^2
   +{g^2\over 2}\left|H^\dagger \widetilde L_I\right|^2\,.
\label{pot}
\eeqn
The soft SUSY breaking parameters are obtained
by numerically solving the renormalization group equations (RGE)
given in Appendix~A with universal boundary conditions [eq.~\ref{softbc}].
For a qualitative understanding we present the result
for $\tanb = 20$ and $m_t = 175~\gev$.
\beqn
B_{I J}        &=& \left(B_0 + 0.25 m_{1/2} - 0.35 A_0\right) \delta_{I J}
                + \Delta B_{I J}\,, \nonumber \\
m_{L_{I J}}^2  &=& \left(m_0^2 + 0.5 m_{1/2}^2\right) \delta_{I J} 
                + \Delta m_{L_{I J}}^2\,,
\eeqn
The non-universal terms are generated through the down-type
Yukawa couplings. For a qualitative understanding
it is convenient to neglect these terms\cite{yossi}.
They are to a good approximation independent of $m_t$
assuming the top Yukawa coupling is not too close to the Landau pole
and have a trivial $\tanb$ dependence
\beqn
\Delta B_{0 0} &\simeq& {10^{-4}\over \cos^2 \beta}
                  \left(2.2 m_{1/2}-1.2 A_0 \right)\,, \nonumber \\
\Delta B_{i j} &\simeq& y^{L*}_{i k} y^L_{j k}
                  \left(0.12 m_{1/2}-0.3 A_0 \right)\,, \nonumber \\
\Delta m_{L_{0 0}}^2 &\simeq&-{10^{-4}\over \cos^2 \beta}
   \left(3.7 m_{0}^2+11.8 m_{1/2}^2+1.0  A^2_0 - 3.6  A_0 m_{1/2}\right)\,,
\nonumber \\
\Delta m_{L_{i j}}^2 &\simeq&-y^{L*}_{i k} y^L_{j k}
   \left(0.9 m_{0}^2+0.25 m_{1/2}^2+0.28 A^2_0 - 0.21 A_0 m_{1/2}\right)\,,
\eeqn
with all other off-diagonal elements vanishing.
We introduce the following notation
for the individual components
\beqn
H = \left(\matrix{H^+\cr (H^0+i A^0)/\sqrt{2}}\right)\,,\qquad
\widetilde L_I = \left(\matrix{(\widetilde L^0_I-i \widetilde B^0_I)
  /\sqrt{2}\cr-\widetilde L^-_I}\right)\,,
\label{def-higgs}
\eeqn
The values of $\bar v$ and $v_I$ are obtained by minimizing the potential
in eq.~\ref{pot} numerically. For small $R$-parity violating
parameters we can also obtain very reliable analytic expressions
in the basis where $y^L_{i j}$ is diagonal
by expanding in powers of $\mu_i/\mu_0$
\beqn
\sin 2 \beta &=& 
         {2 B_{00} \mu_0 \over m_{L_{00}}^2+m_H^2+2 \mu_0^2}\,,\label{sin2b} \\
\tan^2 \beta &=& {m_{L_{00}}^2+ \mu_0^2+\half \mz^2
                 \over m_H^2  + \mu_0^2+\half \mz^2}\,,\label{tanb2}\\
{v_i\over v_0}&=& \mu_i {B_{(i i)} \tanb - \mu_0
                    \over m_{L_{(i i)}}^2+\half \mz^2 \cos 2\beta}\,,
\label{viv0}
\eeqn
with the convention that indices in braces are not summed over.
In order to stay as close to the notation of the MSSM as possible we
define
\beqn
\barv \equiv {\vev{H^0}\over \sqrt2}\,,~
v_I   \equiv {\vev{\widetilde L^0_I} \over \sqrt2}\,,~
v     \equiv  \sqrt{v_I v_I}\,,\hbox{and}  ~
\tanb \equiv \barv/v\,,
\eeqn
and we parameterize the VEVs in terms of spherical
coordinates
\beqn
\tan\theta_3^\prime = {v_3\over v_2}\,,\qquad
\tan\theta_2^\prime = {v_2\over v_1 \cos \theta_3}\,,\qquad
\tan\theta_1^\prime = {v_1\over v_0 \cos \theta_2}\,.
\label{deftani}
\eeqn
Analogously, it is convenient to parameterize the $R$-parity breaking
mass parameters in terms of three mixing angles
\beqn
\tan\theta_3 = {\mu_3\over \mu_2}\,,\qquad
\tan\theta_2 = {\mu_2\over \mu_1 \cos \theta_3}\,,\qquad
\tan\theta_1 = {\mu_1\over \mu_0 \cos \theta_2}\,.
\label{deftaniprime}
\eeqn
and $\mu \equiv \sqrt{\mu_I \mu_I}$.
The potential in eq.~\ref{pot} is minimized by an iterative procedure
using the analytic solution for $\tan \theta_1 = 0$ 
as our initial values.
This procedure also works surprisingly well for $\tan \theta_1 >1$.
However, in order to obtain qualitative understanding of the
results it is instructive to investigate the potential analytically.
Let us for the moment neglect the effects of the
down-type Yukawa couplings on the running of the soft 
SUSY breaking parameters.
Then the conditions
$m_{L_{I J}}^2 \propto \delta_{I J}$ and $B^\mu_I \propto \mu_I$
would continue hold even after renormalization group evolution
down to the electro-weak scale.
Let us  make a rotation on the Higgs/lepton superfields
$\hat L_I \rightarrow   {\cal U}_{I J}^{\dagger}\hat L_J$
with the unitary SU(4) matrix defined by:
\beqn
\mu_I {\cal U}_{I J} = \sqrt{\mu^2} (1, 0, 0, 0)\,.
\eeqn
In this basis there is no $R$-parity violation
in the Higgs potential of the model.
Instead there are $R$-parity violating Yukawa couplings.
For a qualitative understanding this basis is much 
more convenient\cite{suzuki}.
In particular, we find that 
universal soft SUSY breaking parameter at $m_z$ lead to
alignment of $v_I$ and $\mu_I$ which means $\theta_i = \theta_i^\prime$.
It is clear that
by solving the RGEs including the bottom Yukawa coupling,
$y_b\propto m_b \tan \beta$, we have
\beqn
{\theta_1 - \theta_1^\prime\over \theta_1}
\propto
{3 y_b^2\over 16 \pi^2}
\ln{M_{GUT}\over m_Z}
\,.
\label{delta-theta}
\eeqn
This result shows no explicit $\tanb$ dependence
(for fixed Yukawa couplings).
However,  eq.~\ref{delta-theta}
is too naive as we will see by considering the limit
of large $\tanb$. In this case $B_{00}$ vanishes [eq.~\ref{sin2b}]
but $B_{(i i)}$ does not.
This leads to an enhancement of the ratio $v_i/v_0$ [eq.~\ref{viv0}].
In general, we find that eq.~\ref{sin2b}--\ref{viv0}
become unreliable if $\Delta B_{I J} \gsim B_{00}$.
In this case $v_I$ and $\mu_I$ are completely misaligned and
our numerical minimization procedure must be applied.

Let us now turn to the complete particle spectrum of the Higgs sector.
There are
five CP-even neutral scalars,
five CP-odd neutral scalars and eight charged scalars
\beqn
H_x^0 &=& \calu^{H^0}_{x y}(H^0, \widetilde L^0_I)_y
   \qquad (x, y = 1,..,5),\nonumber\\
A_x^0 &=& \calu^{A^0}_{x y}(A^0, \widetilde B^0_I)_y
   \qquad (x, y = 1,..,5),\nonumber\\
H_a^\pm &=& \calu^{H^\pm}_{a b}(H^\pm, \widetilde L_I^\pm, 
\widetilde E_i)_b\qquad (a,b = 1,..,8)\,.
\eeqn
This fields are mixed states of Higgs bosons and sleptons.
The unitary matrices $\calu^\phi$ ($\phi = H^0, A^0, H^\pm$)
are obtained by diagonalizing the corresponding mass matrices
given in Appendix~B.
We define the unitary matrices such that $m_{\phi_i}\leq m_{\phi_j}$
for $i<j$. 
Note, that the CP-odd neutral fields and the charged fields
contain an unphysical mass-less goldstone-bosons, $A^0_1$ and $H^\pm_1$,
which are absorbed via the Higgs mechanism.
These fields cannot occur as external fields 
but they do contribute in loops.

\chapter{Neutrino/Neutralino Sector}

The $R$-parity violating parameters $\mu_i$ are expected to
have the most noticeable effect on the neutrino/neutralino sector.
In models with broken $R$-parity the
neutrinos and neutralinos mix via the mass matrix in eq.~\ref{defyr}.
However, only one of the neutrinos, presumable $\nu_{\tau}$,
acquires a mass at tree-level
contrary to the Majoron models where also $m_{\nu_\mu}$ is 
generated at tree-level\cite{neutrino-c2}.
Furthermore, the neutrino/neutralino mixing turns out to be rather small
in the phenomenologically allowed region. Thus, it is natural to 
keep the conventional MSSM terminology and to
refer to the three lightest (four heaviest) mass eigenstates
as the neutrinos (neutralinos).

In order to obtain the leading contribution to the
masses of the two lightest neutrinos we have to compute
one-loop radiative corrections to the
mass matrix.

\begin{figure}[phbt]
\baselineskip=8pt
\vskip2.5cm
\includegraphics{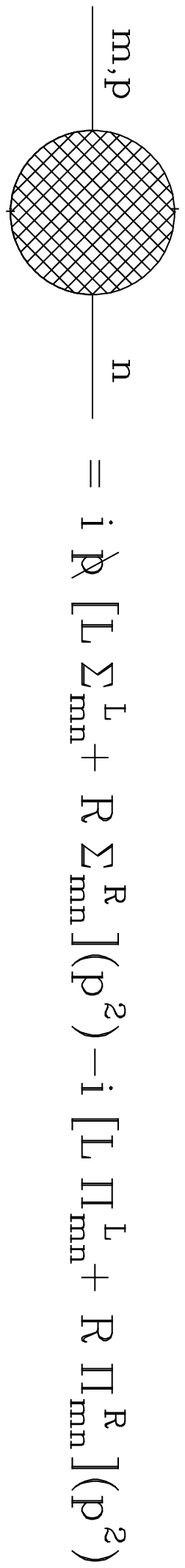}
\caption{\label{fig1}
The Lorentz invariant decomposition of the neutrino/neutralino
self energy.}
\end{figure}

The complete radiative corrections to the neutralino mass
matrix in the MSSM with unbroken $R$-parity
has been presented in ref.~\cite{damien}
and the leading order term for the neutrino masses
has already been derived in ref.~\cite{suzuki}.
The one-loop corrected neutrino/neutralino mass matrix
is given by
\beqn
{\cal M}_{\chi^0_{m n}}^{\rm pole} = 
{\cal M}_{\chi^0_{m n}}^{\drbar}(\mu_R)
+\left[        \Pi^V_{m n}(\barm^2) 
- m_{\chi^0_n} \Sigma^V_{m n}(\barm^2) 
\right]_{\Delta = 0}\,.
\label{drmasses}
\eeqn
Here, the divergences are regularized by dimensional 
reduction\cite{dimred}
where $\mu_R$ is the renormalization scale and $d$ is the
number of space-time dimensions.
The superscript $\drbar$ denotes the renormalized
mass matrix in a minimal subtraction scheme
obtained by setting $\Delta\equiv 2/(4-d)
-\gamma_E+\ln(4\pi)=0$ where $\gamma_{\rm E}$ is the Euler constant.
The Lorentz invariant amplitudes are defined in fig.~\ref{fig1}
and the result for the one-loop self energies are presented in 
Appendix~D\footnote{It was shown in ref~\cite{kniehl} that
 the tadpole diagrams have
to be included for a gauge-independent definition of the 
running masses.
Here, we neglect the tadpole diagrams as they do not 
contribute to $m_{\nu_e}$ and $m_{\nu_\mu}$.}

For the mass eigenvalues it is usually sufficient to consider only the
radiative corrections to the diagonal matrix elements of the propagator
(\ie\ $n= m$). However, in the case of mass-degenerate states
at tree-level such as the two lightest neutrinos
one finds the mass eigenvalues by rediagonalizing
the one-loop corrected neutralino mass matrix.
In addition, we are interested in neutrino oscillations
determined by the mixing angles.
Since we are only interested in the leading effects
it would be sufficient to consider only the one-loop
corrections to the $\nu_e$--$\nu_\mu$ mixing which is 
undetermined at tree-level.
However, for a systematic treatment 
we evaluate the one-loop corrections to all matrix elements and 
rediagonalize the full $7\times7$ mass matrix. 

We define the one-loop mixing matrix, $\Delta Z$
such that the matrix
$m_{\chi^0}= ({\bf 1}+\Delta Z){\cal M}_{\chi^0}^{\rm pole}
({\bf 1}+\Delta Z^{\dagger})$
is diagonal with the 
the one-loop radiatively corrected neutrino/neutralino masses
$m_{\chi_n^0}$ as the diagonal elements.
The choice for the momentum at which the
off-diagonal elements have to be evaluated is ambiguous.
However, the effects of this ambiguity is of higher order since the 
off-diagonal elements are only important if $m_{\chi_n^0}-m_{\chi_m^0}
\lsim  O(\alpha/4\pi)$ and 
we simply choose $p^2 = \barm^2 \equiv (m_{\chi_n^0}^2
+m_{\chi_m^0}^2)/2$

In a complete physical scheme any explicit dependence on 
gauge parameter or
renormalization scale should cancel.
A complete analysis would require
the renormalization of the parameters $v, g, g^\prime, \tanb, \mu, M$
and $M^\prime$ at one-loop.
However, we are only interested in the leading contributions
to the neutrino masses and mixings. 
Thus, the one-loop corrections are only relevant for the two 
lightest neutrino
masses. We have checked that the radiatively generated
masses of the lightest two neutrinos are indeed gauge independent
and renormalization scale independent.
For the third neutrino (which acquires mass at tree-level) the
scheme-dependence is of higher order and can be neglected.

\section{The neutrino mass spectrum: A general Scan}

In order to see what a typical neutrino mass 
spectrum looks like we first perform a general scan
over the entire parameter space. We impose the standard assumption
that all the soft SUSY breaking parameters are universal at 
$\mgut$. The spectrum then depends on four
SUSY parameters, $\tanb$, $A_0$, $m_{1/2}$ and $\mu$
(here, the value of $m_0$ is 
determined by imposing radiative electro-weak symmetry breaking and $B$
is replaced in favor of $\tanb$) and in addition on three $R$-parity
violating angles $\tan\theta_i$ ($i=1,2,3)$.

From our Appendix~C we see that the lepton current coupling to the $W^-$
can be written as
\beqn
J_\mu^+ = {\sqrt{2}\over g} {\overline\chi}^0_n \gamma_\mu 
\left(L O^L_{n x} + R O^R_{n x}\right)\chi_x^+\,.
\eeqn
Here, $\sqrt{2}O_{n x}^R/g$ is the analog to the CKM 
matrix of the quark sector.
Note that in our notation the conjugate left-handed lepton doublets
form the right-handed component of the charginos and vice versa.
This unconventional arrangement was neccessary in order to 
have a unified notation for neutrinos and neutralinos
and for charged leptons and charginos.

The one-loop radiatively corrected interaction marices
$O^L$ and $O^R$ are obtained from eq.~\ref{rotations}
by replacing $Z\rightarrow ({\bf 1}+\Delta Z) Z$.
The analogous one-loop corrections to the charged lepton
mixing matrix is suppressed by the inverse power of the lepton masses
and can savely be neglected.

There are two fundamental difference to the quark sector:
(i) the existence of a left-handed and a right-handed current
and (ii) the matrices $\sqrt{2}O_{n x}^P/g$ ($P = L,R$) are not unitary.
However, it turns out that only the case
$\tan \theta_1 \ll 1$ is phenomenologically interesting.
In this case, the $3\times3$ submatrix $O_{i j}^L \ll 1$ 
($i,j = 1,2,3$)
can be neglected. The $3\times3$ submatrix $O_{i j}^R$
is to a very good approximation unitary
and equivalent to the leptonic CKM matrix. 
Furthermore, in order to adopt a more standard notation we
define
\beqn
\sin \theta_{\ell_i \nu_{j}} \equiv {\sqrt{2}\over g}
 \abs{O_{i j}^R}\,,
\qquad i\neq j\,.
\eeqn
Strictly speaking, this is only correct in the
case of two generation mixing
or for small mixing angles, but the accuracy
is still better than the experimental uncertainties and 
will be sufficient for our purposes.

It is somewhat problematic to present the result of different diagrams
given in eq.~\ref{defp} and \ref{defs}
because the mass eigenvalues of a sum of two matrices is not
equal to sum of mass eigenvalues of two matrices.
Thus we will only consider the full result obtained by
summing over all one-loop diagrams. However, we would like to emphasize
that the only significant diagrams are the ones envolving the down-type
quarks/squarks and the charginos/charged Higgs bosons.
It turns out that for small values of $\tanb$ the down-type quark/squark loops
dominate due to the large Yukawa couplings and a color enhancement.
For large $\tanb$ there are large sneutrino VEVs even in the basis 
where $\mu_I \propto (1,0,0,0)$. Here the dominant effects arise from
the chargino/charged Higgs loop with sneutrino VEV insertions
rather than from $R$-parity violating Yukawa couplings.

\begin{figure}
\vspace*{13pt}
\vspace*{6.6truein}        
\includegraphics{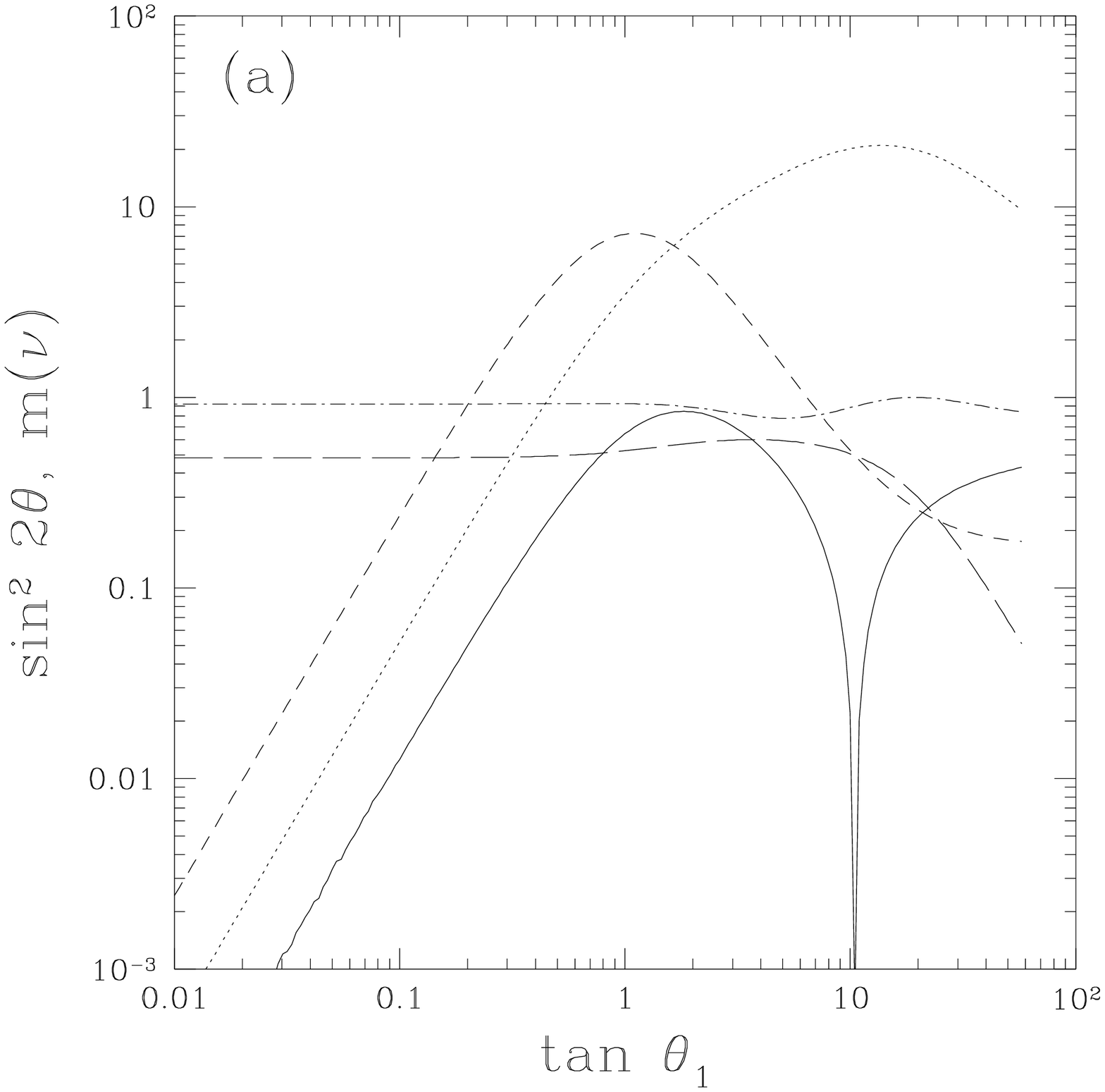}
\includegraphics{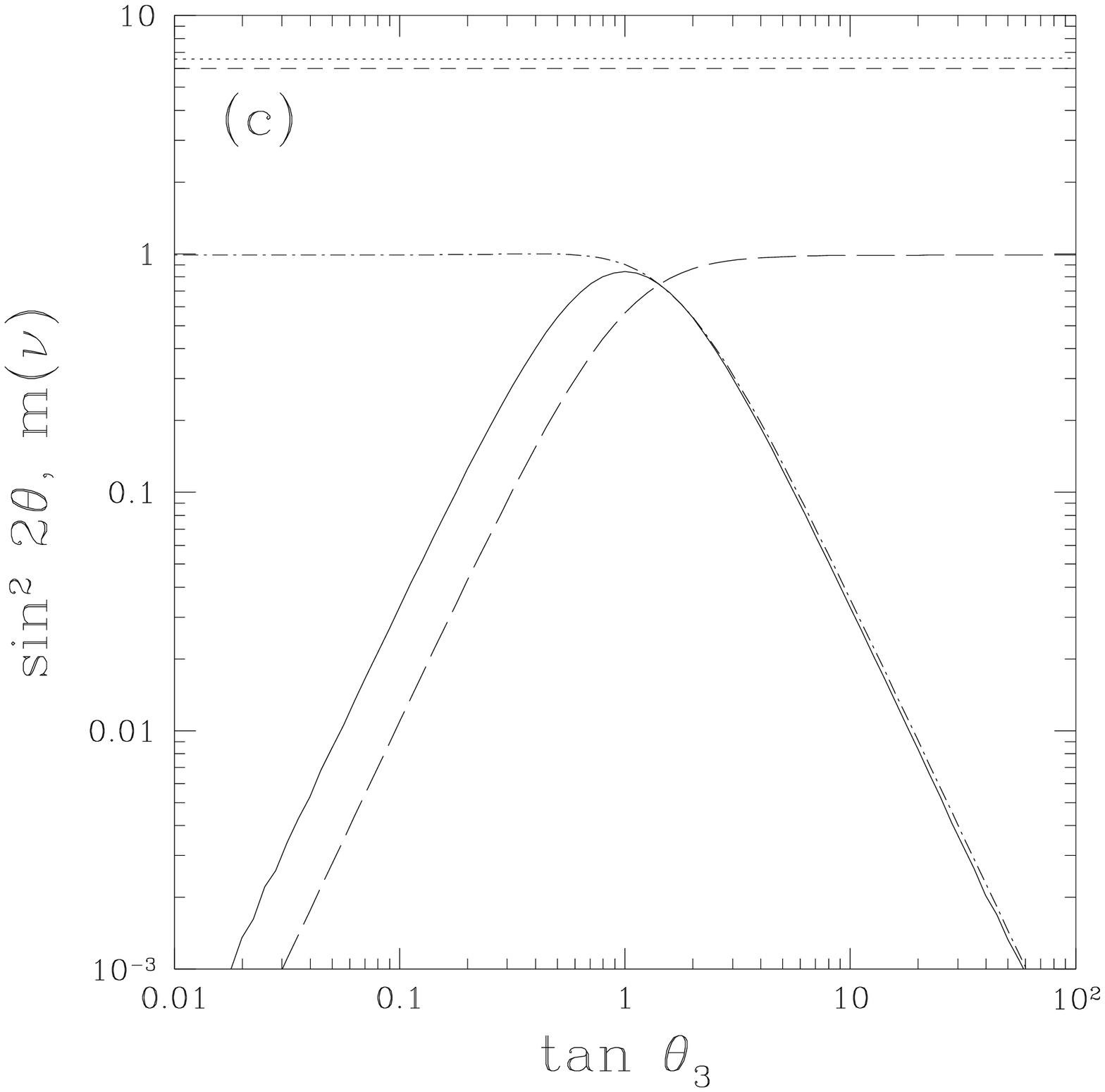}
\includegraphics{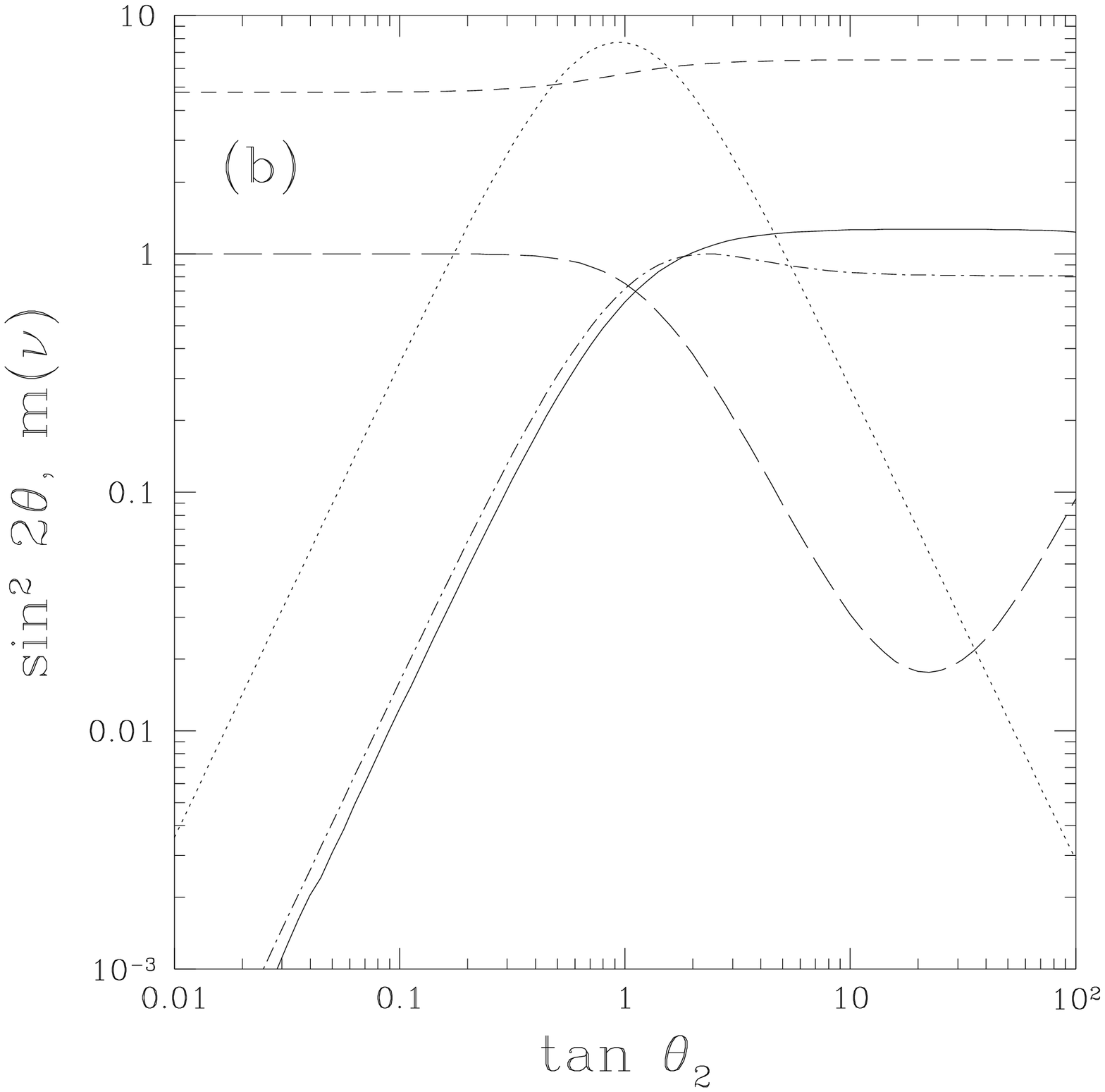}
\includegraphics{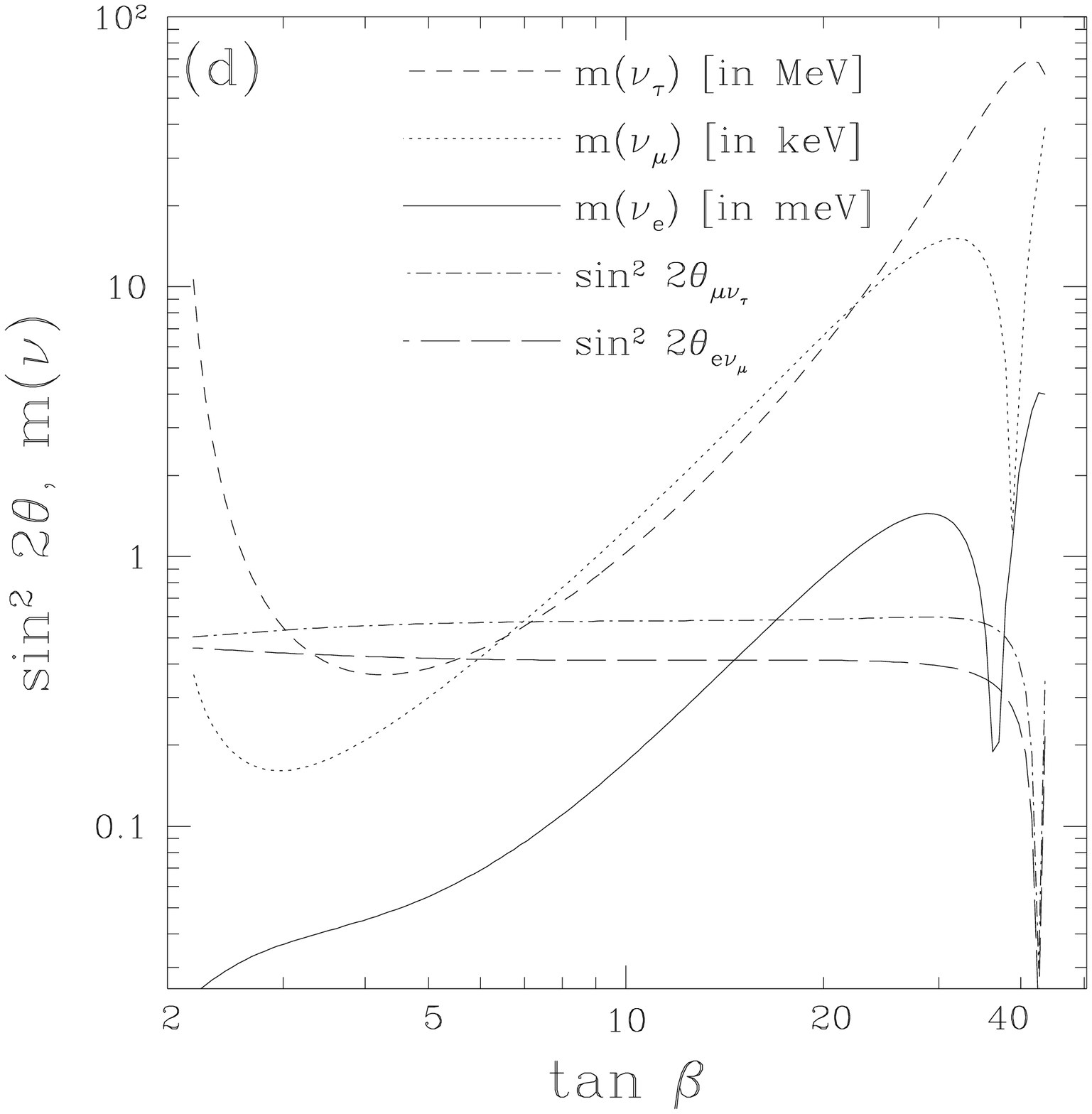}
\caption{The one-loop radiatively corrected neutrino masses
as a function of
(a) $\tan \theta_1$, (b) $\tan \theta_2$, (c) $\tan \theta_3$ and
(d) $\tan \beta$. Our master set of parameters is
$\mu = 200~\gev$, $A_0 = 0$, $m_{1/2} = 100~\gev$
and we chose $\tan^2 \theta_1 = 3$, $\tan^2 \theta_2 = 2$,
$\tan^2 \theta_3 = 1$,
$\tan \beta = 20$ and $m_t = 175~\gev$
 whenever these variables are not varied.}
\label{fig2}
\end{figure}

\begin{figure}
\vspace*{13pt}
\vspace*{6.6truein}      
\includegraphics{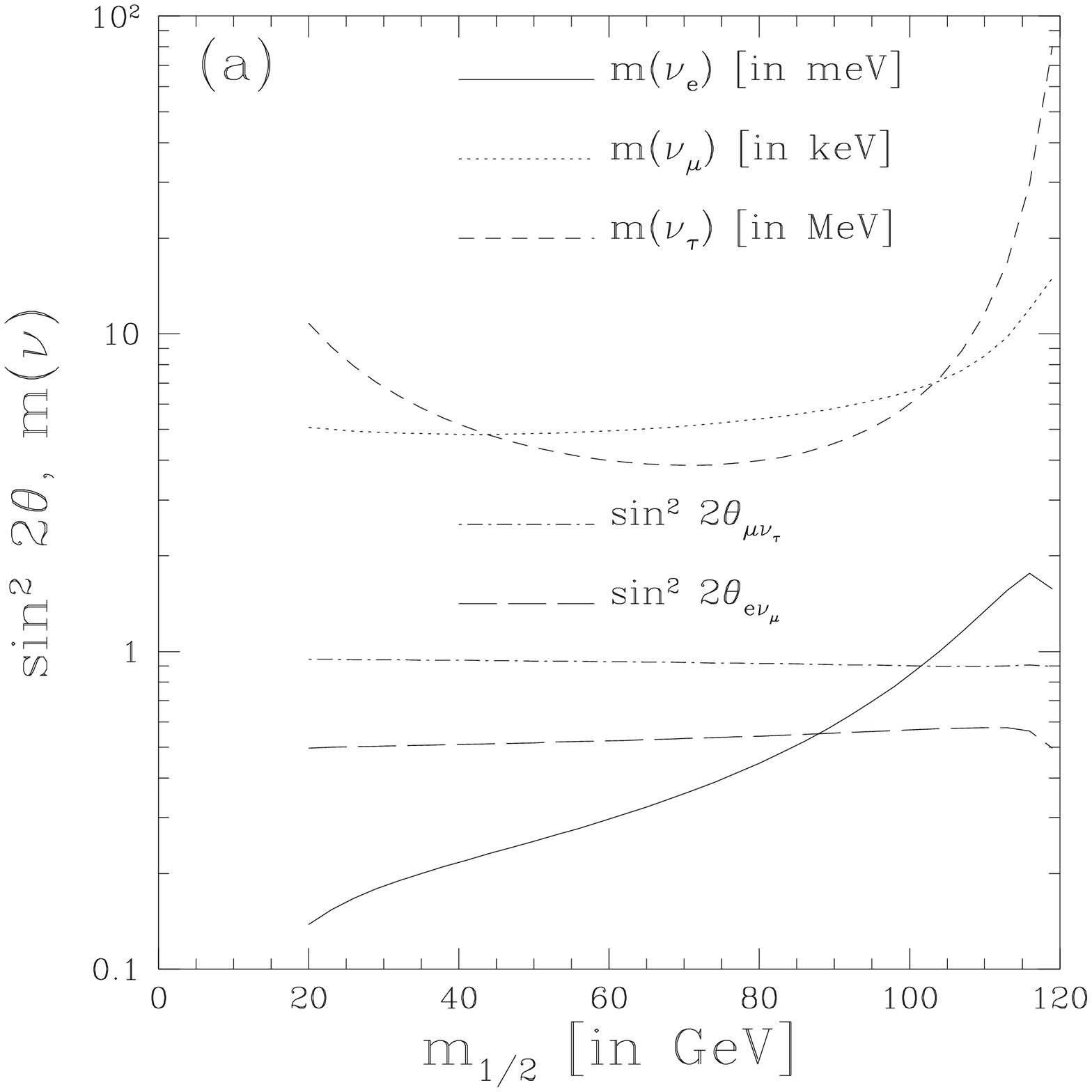}
\includegraphics{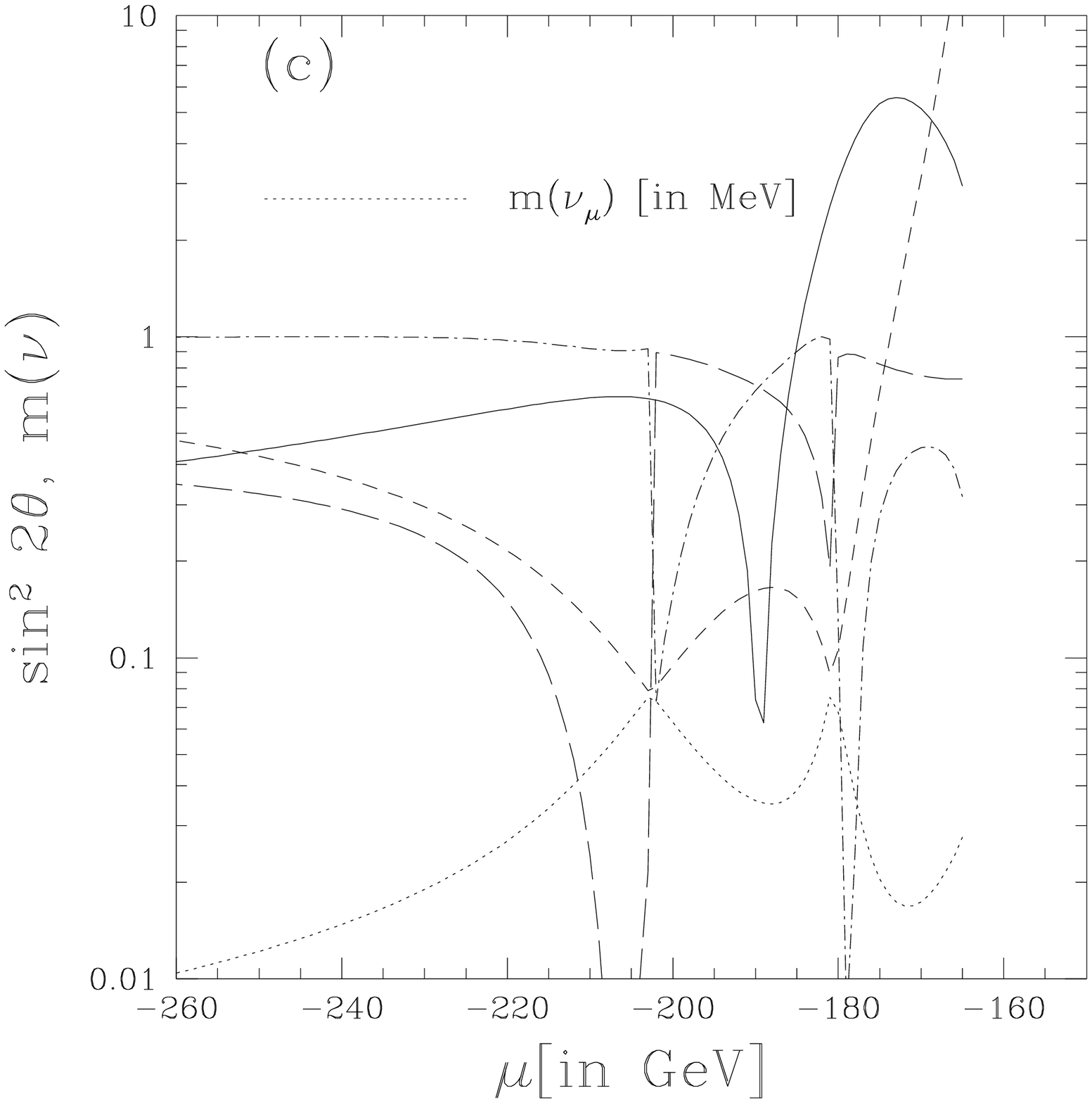}
\includegraphics{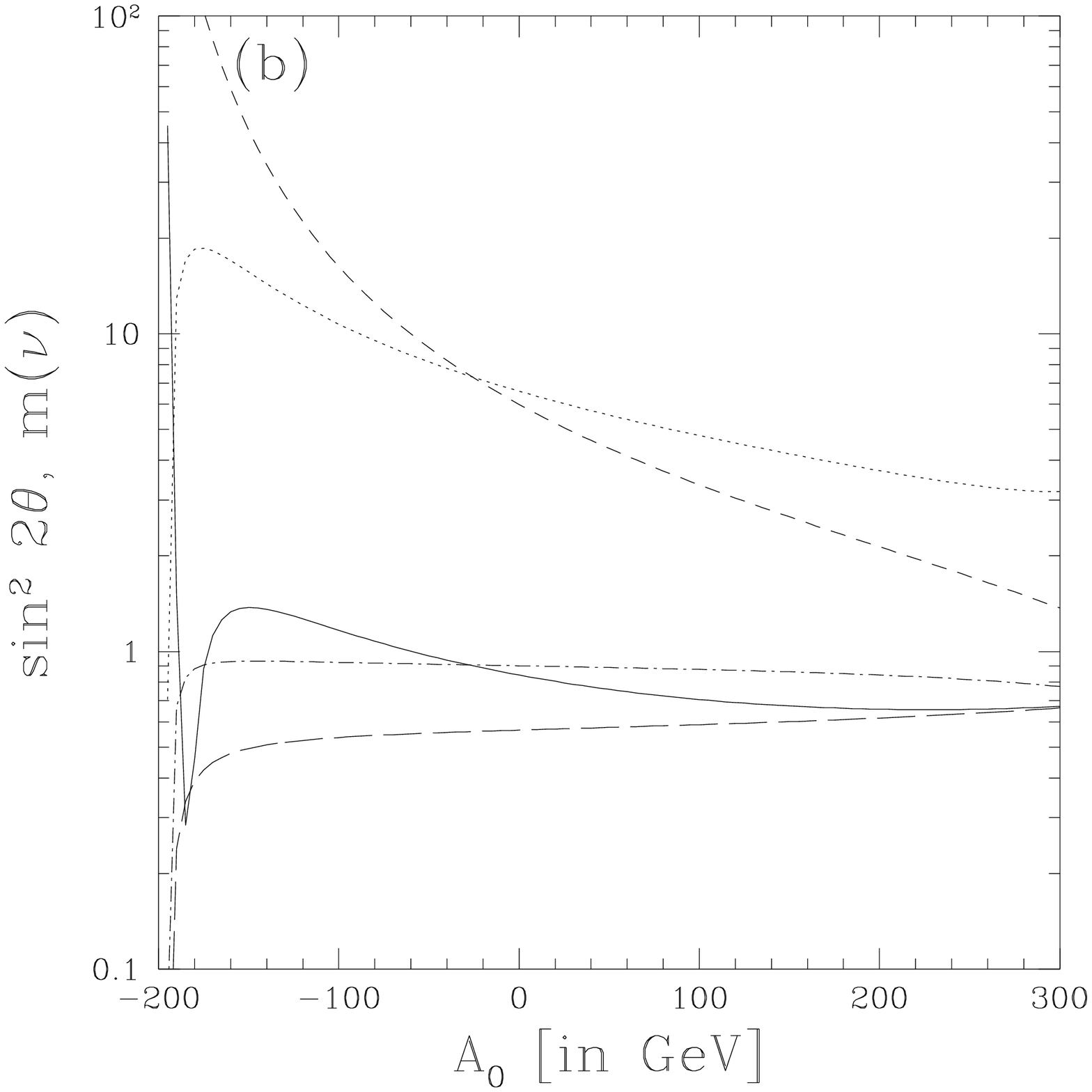}
\includegraphics{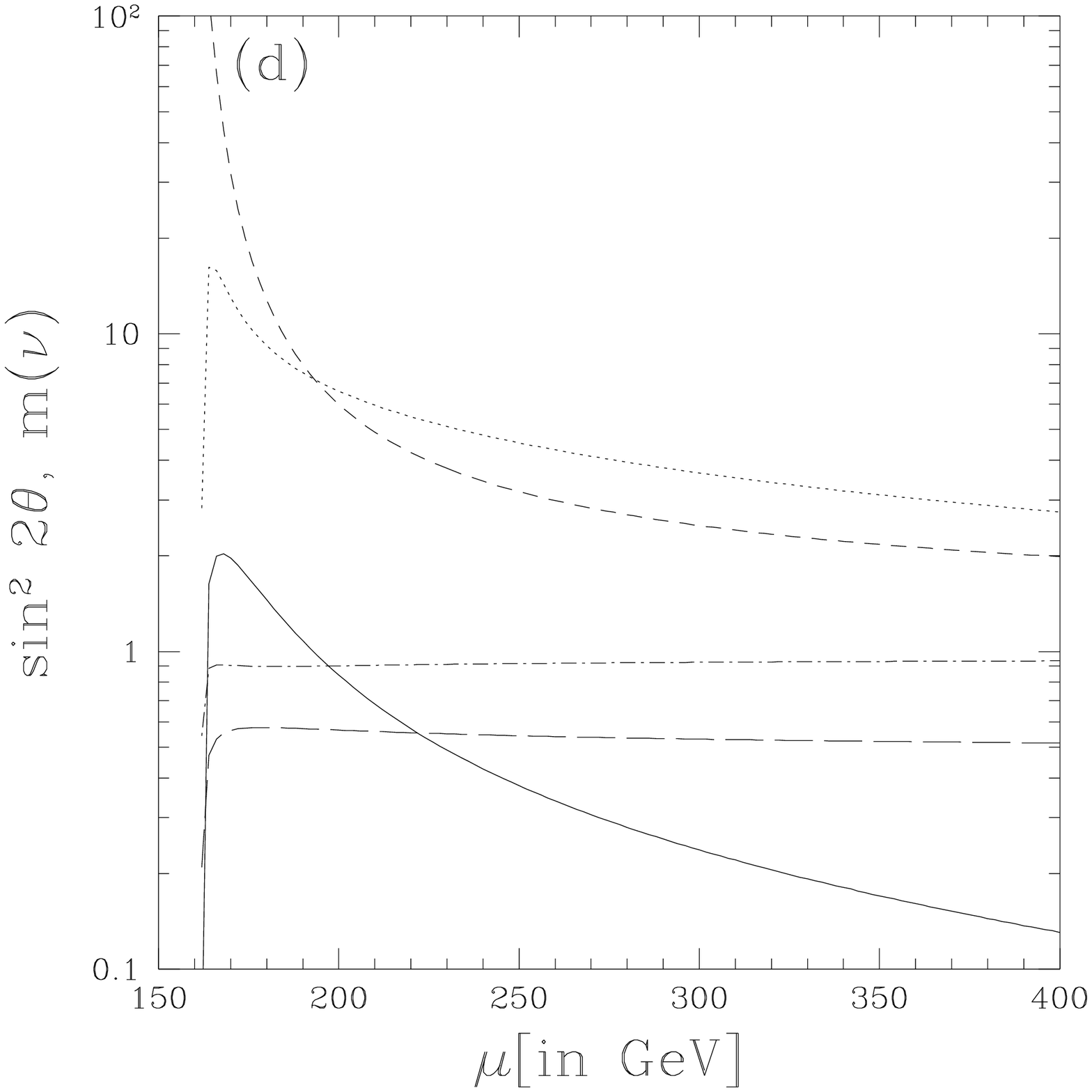}
\caption{The one-loop radiatively corrected neutrino masses
as a function of
(a) $m_{1/2}$, (b) $A_0$, (c) $\mu<0$ and
(d) $\mu>0$.
Our master set of parameters is as in fig.~\ref{fig2}.}
\label{fig3}
\end{figure}

In fig.~\ref{fig2} and \ref{fig3} we show the dependence of
the three neutrino masses and the $e$--$\mu$ and 
$\mu$--$\tau$ mixing angles as a function of all seven parameters.
Our master set of parameters is:
$A_0 = 0$, $m_{1/2} =100~\gev$, $\mu = 200~\gev$, $\tanb = 20$
and we assume maximum mixing, \ie\ $\mu_I = \half \mu(1,1,1,1)_I$
or $\tan \theta_1 = \sqrt3$, $\tan \theta_2 = \sqrt2$,
$\tan \theta_3 = 1$.
Furthermore, we set $m_t = 175~\gev$ in all plots.

In fig.~\ref{fig2} (a)--(c) we see that
\beqn
m_{\nu_\ell}\propto \tan^2 \theta_i\,,\qquad \hbox{where} 
\cases{
i = 1,2,3  & for  $\ell =  e\,,  $\cr
i = 1,2    & for  $\ell = \mu\,, $\cr
i = 1      & for  $\ell = \tau\,,$}\label{mprop}
\eeqn
and
\beqn
\sin^2 2\theta_{e   \nu_\mu } \propto \tan^2 \theta_3\,,\nonumber\\
\sin^2 2\theta_{\mu \nu_\tau} \propto \tan^2 \theta_2\,.
\label{aprop}\eeqn
as long as $\tan \theta_i \ll 1$.
The dependence on $\tanb$ is more complicated.
We see that there is a sharp rise in $m_{\nu_\tau}$ 
if the top Yukawa coupling is near a Landau pole
(\ie\ $\tanb \simeq 2$).
On the other hand, there is also an increase
of all three neutrino masses with $\tanb$.
Note that the explicit $\cos^2 \beta$ suppression
of $m_{\nu_\tau}$ in eq.~\ref{mtau} is overcompensated 
by the $\tan^4 \beta$ dependence of $\sin^2 (\theta_1-\theta_1^\prime)$.

\begin{figure}
\vspace*{13pt}
\vspace*{3.5truein}             
\includegraphics{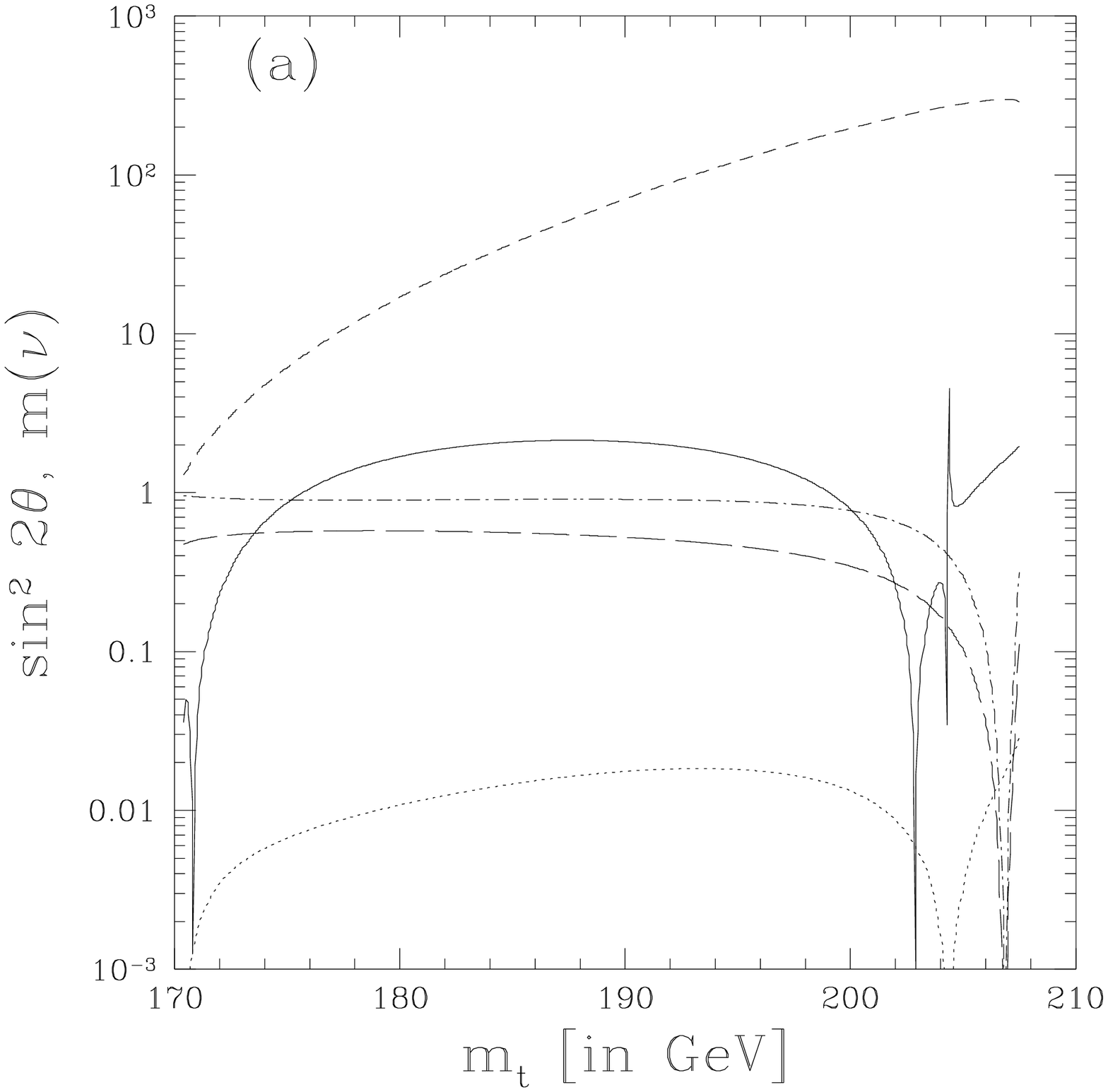}
\includegraphics{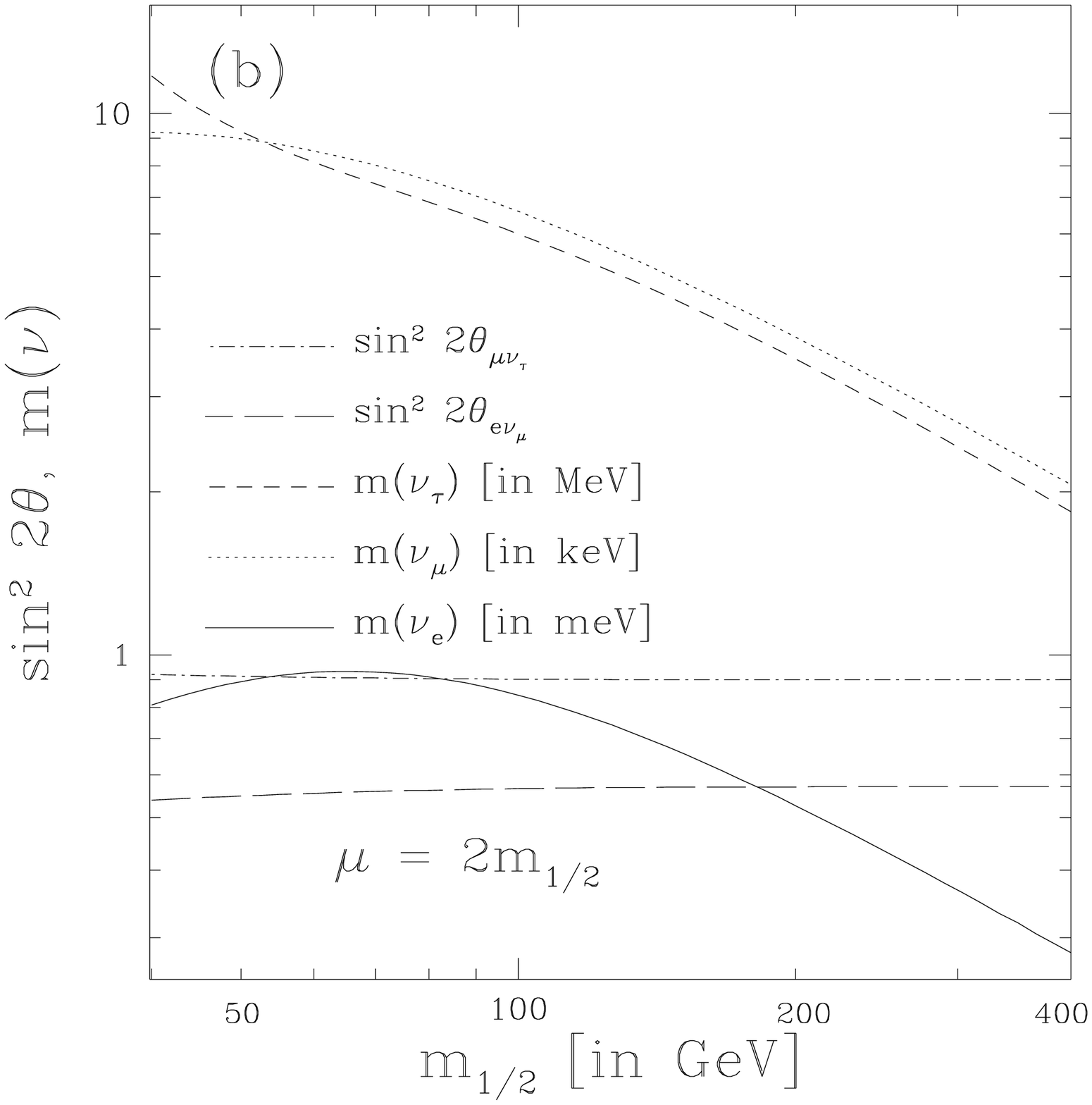}
\caption{The one-loop radiatively corrected neutrino masses
as a function of (a) the top quark mass and
(b) the SUSY breaking scale parameterized by $m_{1/2}$.
We have set $A = 0$ and $\mu = 2 m_{1/2}$.}
\label{fig4}
\end{figure}

In fig.~\ref{fig2}(d) no obvious proportionality 
can be established. We can see that
a destructive interference of various
one-loop diagrams can result in a vanishing
$m_{\nu_\mu}$ for a particular (large) value of $\tanb$.
The same can happen to
$m_{\nu_e}$ for some values of the SUSY parameters.

We find that over most of the parameter space
there is a hierarchy
among the three neutrino masses. However, there are 
regions where the heaviest one-loop
generated mass can dominate over the tree-level mass.
E.g. in fig.~\ref{fig3}(c)
we see that $m_{\nu_{\mu}}$ and $m_{\nu_{\tau}}$ intersect
 for $\mu \simeq -300$
and $-200~\gev$. On the other hand, we find that the mass of the 
lightest neutrino $m_{\nu_e}\ll
m_{\nu_{\mu}}, m_{\nu_{\tau}}$ over the entire parameter space.

In fig.~\ref{fig4} we  present the masses and mixing angles as a
function of the SUSY breaking scale parameterized by $m_{1/2}$
for $A_0=0$ and $\mu = 2 m_{1/2}$.
We find that all the masses are inversely proportional to
the SUSY breaking scale (aside from singularities in the lightest
mass eigenvalue due to some accidental cancellation among
different diagrams). This decoupling is due to the fact that
our model reduces to the SM in the limit of large SUSY breaking scales.

\begin{figure}
\vspace*{13pt}
\vspace*{6.6truein}             
\includegraphics{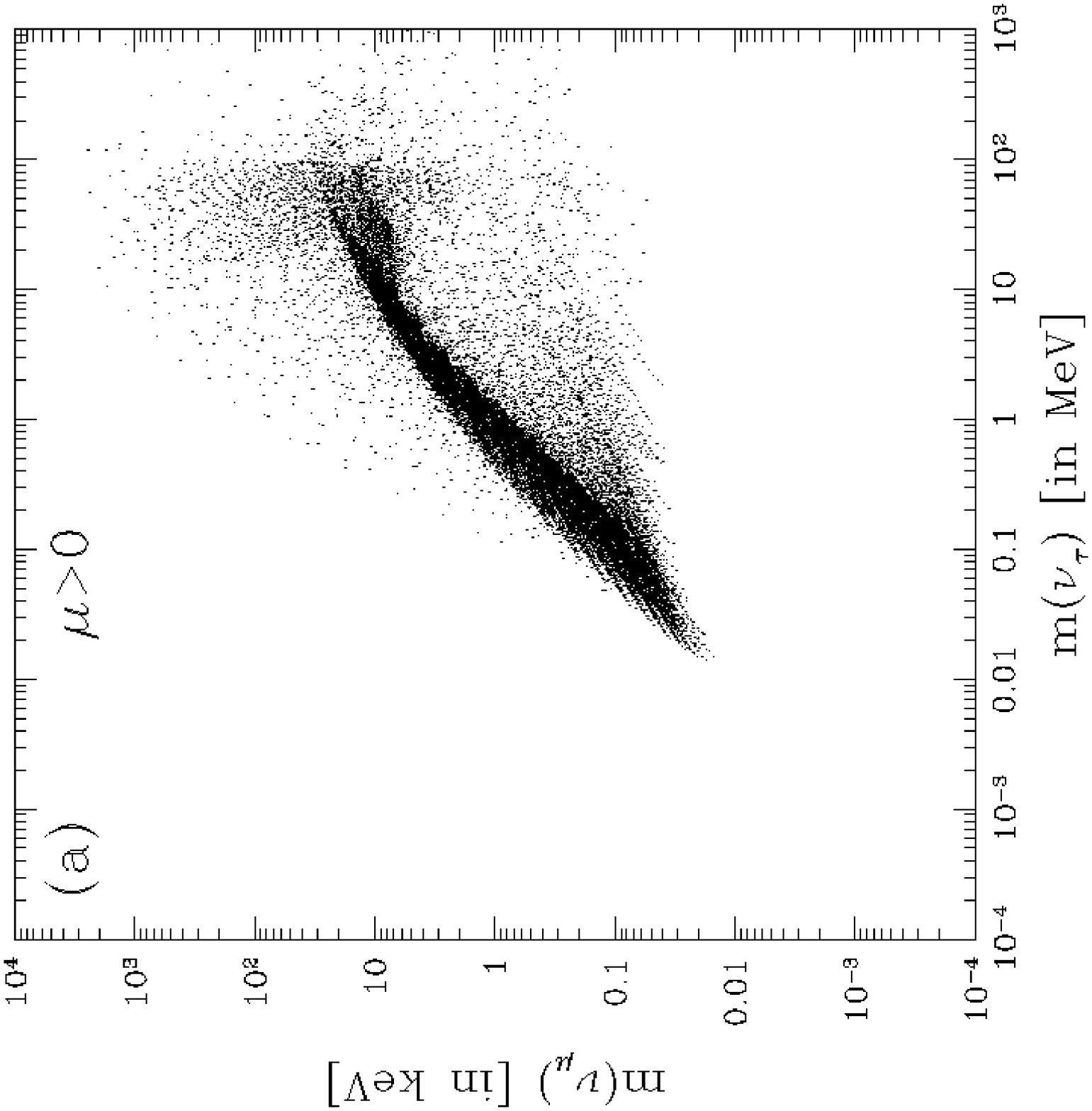}
\includegraphics{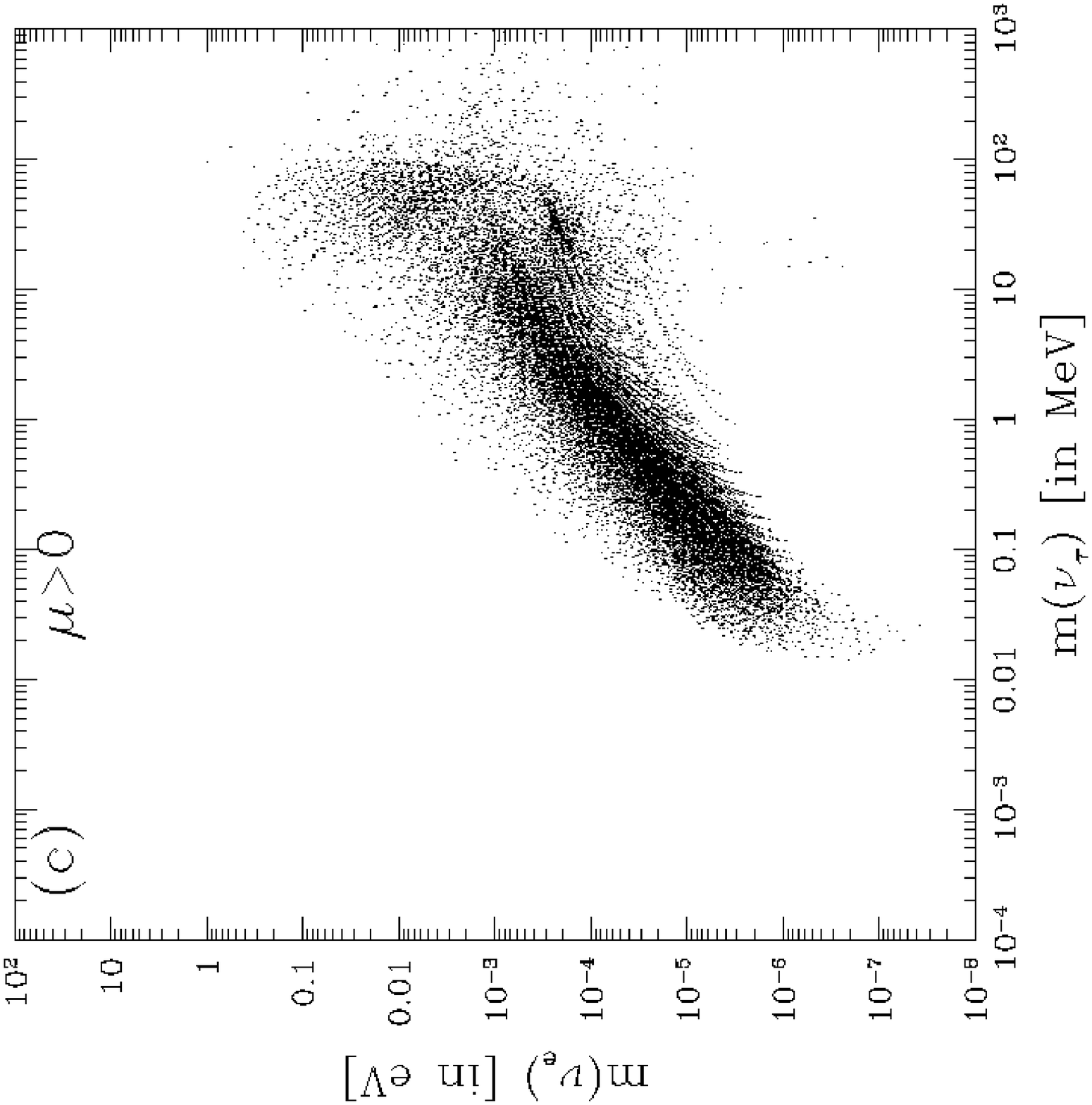}
\includegraphics{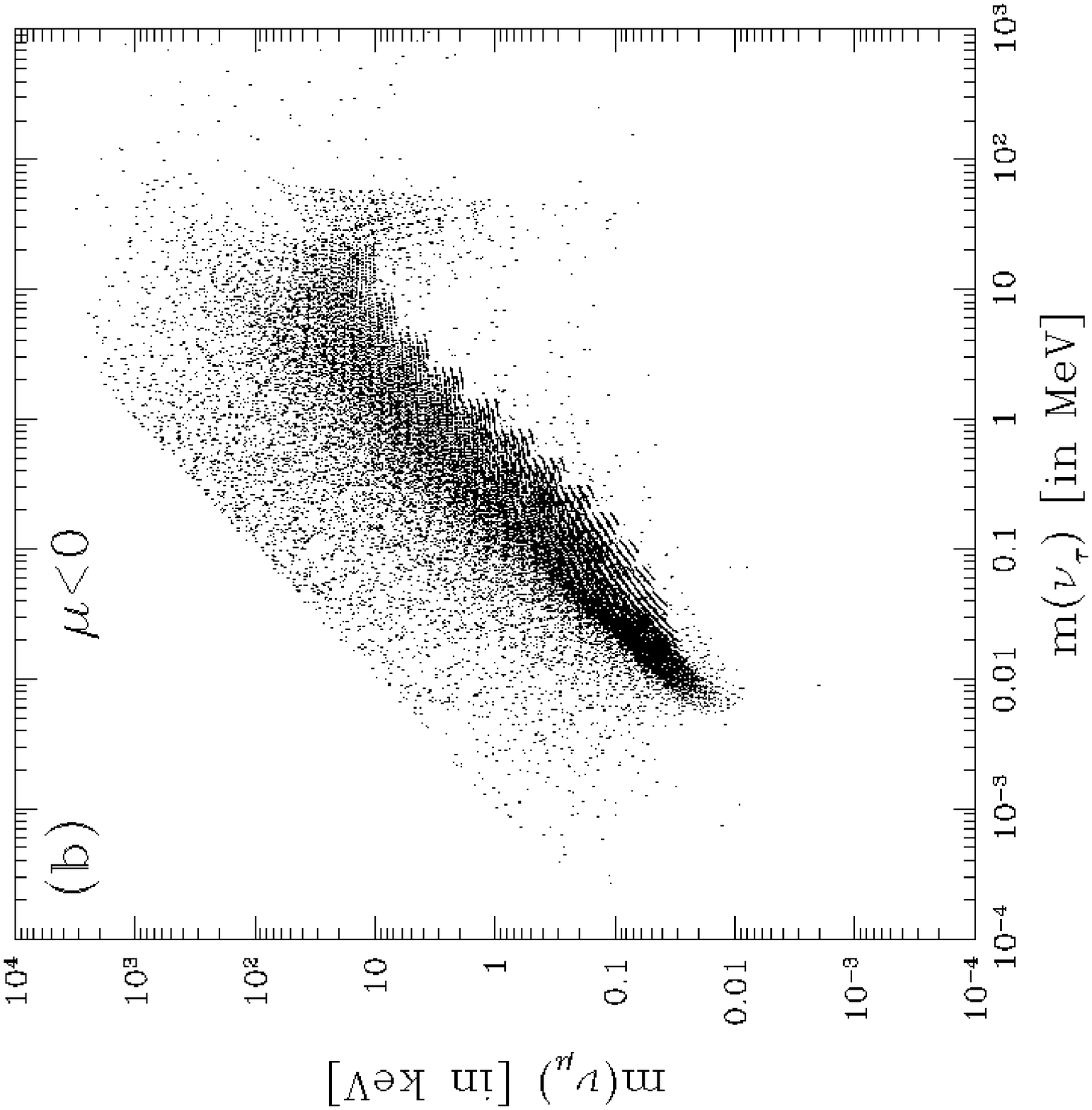}
\includegraphics{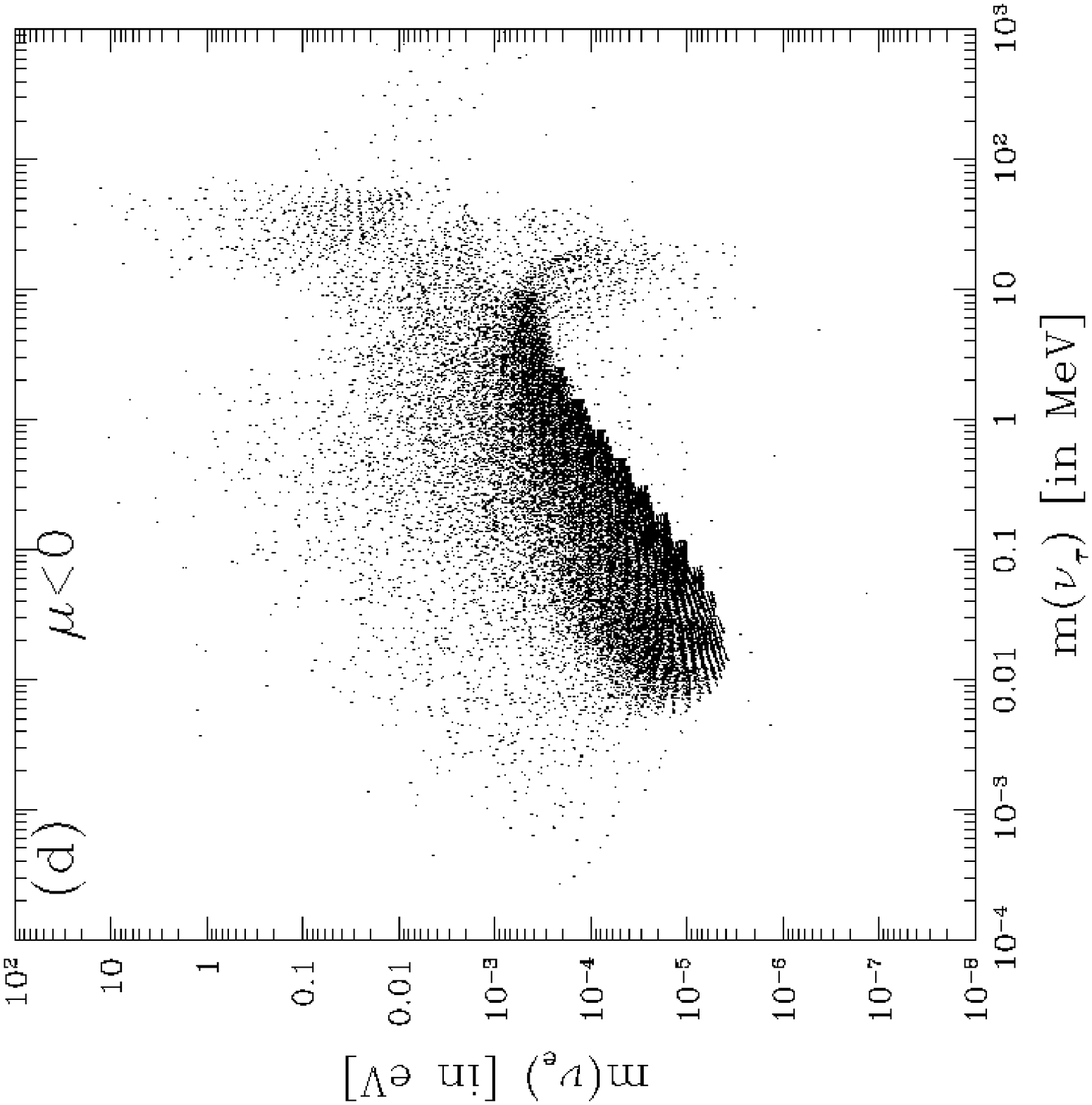}
\caption{
Scatter-plot of $6\times 10^{4}$ models with maximal mixing in the
(a) the $m_{\nu_\tau}$--$m_{\nu_\mu}$ plane and
(b) the $m_{\nu_\tau}$--$m_{\nu_e}$ plane.}
\label{fig5}
\end{figure}

Finally, in fig.~\ref{fig5} we present
scatter plots of $6\times 10^{4}$ different sets of parameters.
Again we assume maximum mixing and we
have scanned over all SUSY breaking parameters in the 
range $2<\tanb<40$, $|A/m_{1/2}|<3$,
$50~\gev\leq m_{1/2}, \abs{\mu} \leq 500~\gev$ and $1<|\mu/m_{1/2}|<5$.
Typically, we find
$m_{\nu_{\mu}}/m_{\nu_{\tau}}\gsim 10^{-4}$ and
$m_{\nu_e}/m_{\nu_{\tau}}\lsim 10^{-7}$.
The fact that the heavier one-loop radiatively generated mass
and the tree-level mass
are relatively close in magnitude can be understood from eqs.~\ref{delta-theta}
and \ref{mtau}:
\beqn
{m_{\nu_\mu}\over m_{\nu_\tau}} = 
{\theta_1\over \theta_1-\theta_1^\prime}
\times
O\left({\alpha_{em}\over 4\pi} \right)\,.
\eeqn
We see that the one-loop suppression is compensated by the smallness
of $\theta_1-\theta_1^\prime$.
Furthermore, we see that  even in the case of maximum $R$-parity violation
the neutrino masses are below their experimental bounds\cite{nmasses}
over most of the parameter space.
The reasons for this suppression was discussed 
in detail in ref.~\cite{yossi}

A neutrino mass spectrum for different mixing 
angles can be obtained to a very good approximation
by using the proportionality
relations of eq.~\ref{mprop} and \ref{aprop}.

\section{Solar and atmospheric Neutrino puzzles}

So far we have only considered the theoretically predicted
neutrino spectrum. We have focused on the case of maximum mixing,
from which
the general case can be derived by rescaling the masses by 
appropriate powers of $\tan\theta_i$ ($i = 1,2,3$).

Before discussing any virtue of a given model we should assure
that it satisfies all experimental bounds.
There are various constraints on the SUSY spectrum.
In particular, from LEP-1 we know that
no superpartner (except a gaugino-like neutrino)
can exist with mass below $\mz/2$\cite{pdg}.
Stronger constraints on strongly interacting sparticle masses
are obtained from experiments at FERMILAB\cite{pdg}.
All of these bounds can be evaded by scaling up the soft
SUSY mass parameters just like in the models with 
unbroken $R$-parity.
Typically any model with $m_0, m_{1/2} \gsim 100~\gev$
provides a sufficiently heavy sparticle spectrum.
One can try to derive stronger constrains from 
virtual effects in processes as $b \rightarrow s \gamma$,
but it requires very strong assumptions on the squark mass 
matrices and possible cancellations among different contributions makes
it hard to derive firm bounds that are much higher
than those coming from direct particle searches\cite{francesca}.

In order to impose the experimental constraints on the Yukawa couplings
on our model we have to rewrite the Lagrangian in a basis
where the Higgs fields are the only ones with non-zero VEV.
In this basis, there exist $R$-parity violating Yukawa couplings
proportional to the lepton masses
$y_{i j k}^F \simeq y_{j k}^F v_i/v_0$ ($F = L, D$).
However, they are below the experimental bounds
even in the case of maximal mixing.

In principle we could now proceed to impose bounds on
neutrino masses\cite{nmasses}
\beqn
m_{\nu_e} &\lsim& 4.35~\ev\,,\nonumber\\
m_{\nu_{\mu}}& \lsim& 160~\kev\,,\nonumber\\
m_{\nu_{\tau}}& \lsim& 23~\Mev\,.
\eeqn
and lepton mixings from the partial $Z$ width\cite{pdg}
\beqn
\Gamma(Z\rightarrow e^\pm \mu^\mp   ) 
&<& 0.6\times 10^{-5}\,,\nonumber\\ 
\Gamma(Z\rightarrow e^\pm \tau^\mp  ) 
&<& 1.3\times 10^{-6}\,,\nonumber\\ 
\Gamma(Z\rightarrow \mu^\pm \tau^\mp) &<& 1.9\times 10^{-6}\,,
\eeqn
in order to obtain constraints on the $R$-parity
breaking parameters\footnote{
A much stronger constraint of $m_{\nu_{\tau}} \lsim 100~\ev$
can be obtained by requiring that the neutrino density
does not overclose the universe\cite{k&t}.}.
This route has been taken by various other authors in models
with spontaneous $R$-parity breaking
or in models with $R$-parity
breaking Yukawa couplings\cite{neutrino-c1}
and very recently also in the model under investigation
 here\cite{pilaftsis}.

However, our goal is slightly more ambitious.
Rather that trying to rule out
a certain region of the seven dimensional parameter space
we are more interested to see
whether our predictive model
with only three $R$-parity violating parameters
can provide the solution to actual problems.
In particular, we want to find out
whether the pattern of neutrino
masses can provide a natural framework for a solution 
of the solar\cite{solarn} and atmospheric\cite{atmosphericn} 
neutrino puzzles.
The masses and mixing angles needed to solve the atmospheric
neutrino puzzle are\cite{atmosphericn}
\beqn
m_{\nu_{\tau}}^2 - m_{\nu_{\mu}}^2\simeq 10^{-2}~\ev^2&\,,&
    \qquad  \sin^2 2\theta_{\mu \nu_\tau} \simeq 1\,.
\label{dtau-mu}
\eeqn
For the solar neutrino puzzle there are three possible regions 
in parameter space\cite{solarn},\cite{lwo}
\beqn
m_{\nu_{\mu}}^2 - m_{\nu_{e}}^2\simeq 5\times 10^{-6}~\ev^2&\,,&
    \qquad  \sin^2 2\theta_{e \nu_\mu} \simeq 0.008\nonumber\\
m_{\nu_{\mu}}^2 - m_{\nu_{e}}^2\simeq 5\times 10^{-6}~\ev^2&\,,&
    \qquad  \sin^2 2\theta_{e \nu_\mu} \simeq 1\nonumber\\
m_{\nu_{\mu}}^2 - m_{\nu_{e}}^2\simeq 10^{-10}~\ev^2&\,,&
    \qquad  \sin^2 2\theta_{e \nu_\mu} \simeq 1\,.
\eeqn
The first two regions correspond to an MSW solution\cite{msw-effect}
and the third one corresponds to long-wavelength oscillations
(LWO)\cite{lwo}.
Our analysis is similar to the 
one performed in ref.~\citenum{neutrino-c2}
and \citenum{rspontaneous2} in models with
spontaneous $R$-parity breaking
and in ref.~\cite{ryukawa}
for models with $R$-parity breaking Yukawa couplings.
The model under consideration here 
distinguishes itself from the later models by permitting
only three $R$-parity violating parameters in addition
to the usual minimal SUGRA parameters and is thus considerably
more constrained. On the other hand, it does not predict any
new particles below the $Z$ mass and cannot easily be tested 
in present\cite{pilaftsis} or future collider experiments.

The desired values of $m_{\nu_{\tau}}^2- m_{\nu_{\mu}}^2$,
$\sin^2 2\theta_{\mu \nu_\tau}$
and $\sin^2 2\theta_{e \nu_\mu}$ can be obtained by
adjusting $\tan \theta_1$, $\tan \theta_2$ and $\tan \theta_3$,
 respectively.
Clearly, the desired value of $m_{\nu_\tau}$ [eq.~\ref{dtau-mu}]
is some four to ten orders of magnitude smaller than the ones 
obtained for maximum
$R$-parity violation [fig.~\ref{fig5}] implying
that $\tan\theta_1$ has to be
rescaled by some two to five orders of magnitude. 
Similarly we can fix $\sin^2 2\theta_{\mu \nu_\tau}$
and $\sin^2 2\theta_{e \nu_\mu}$ by tuning
$\tan \theta_2$ and $\tan \theta_3$.
The value of $m_{\nu_{\mu}}^2- m_{\nu_{e}}^2$ is then predicted
as a function of the SUSY parameters. 
Note that the source of the lepton number violations
and lepton flavor violations
are soft mixing terms that are uncorrelated
with the Yukawa couplings. Thus, there is no
relation between the neutrino mixing angles and the CKM-matrix
of the quark sector
such as in SO(10) SUSY-GUT models and we should forget
any prejudice towards small mixing angles.

\begin{figure}
\vspace*{13pt}
\vspace*{3.2truein}             
\includegraphics{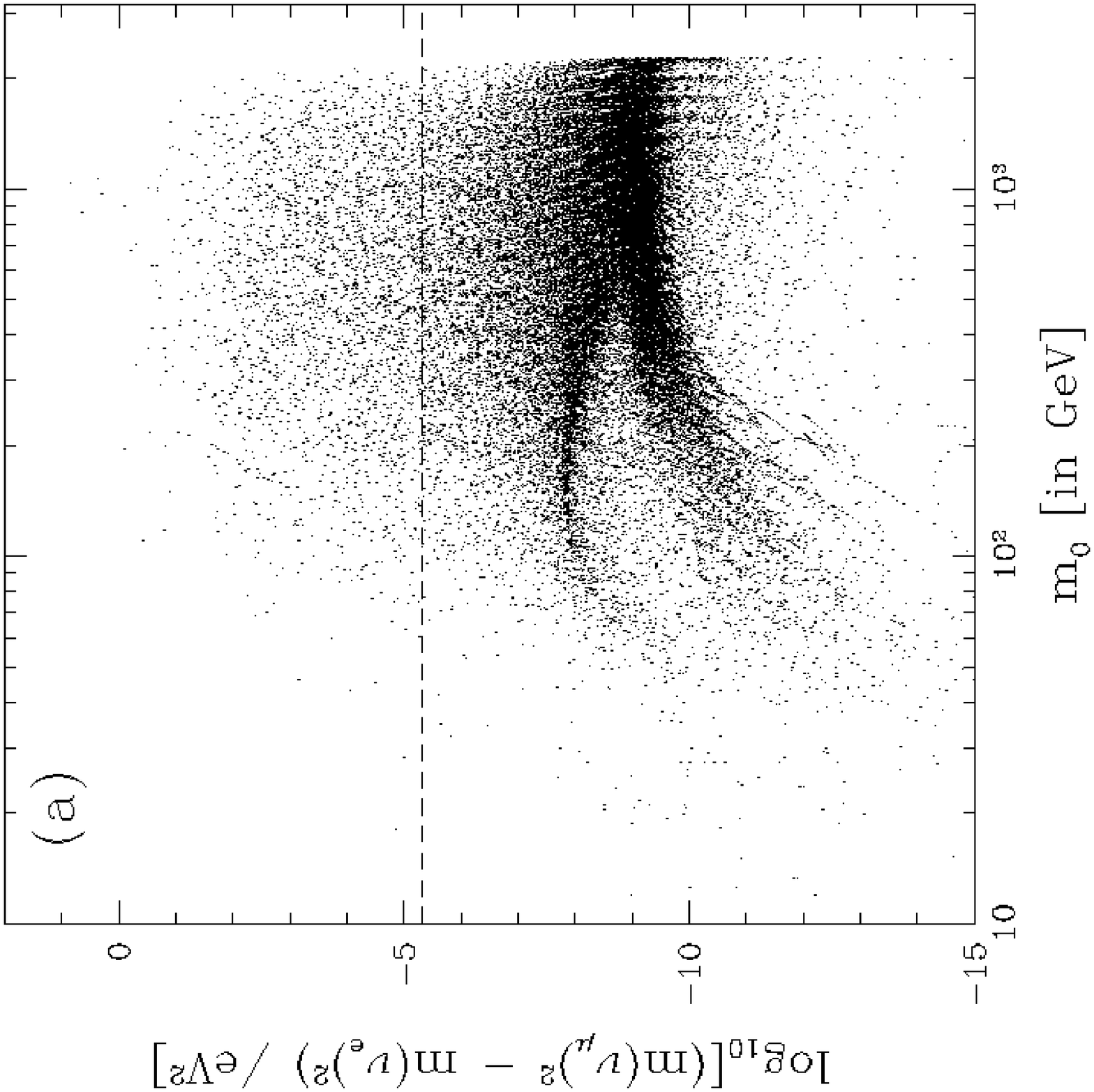}
\includegraphics{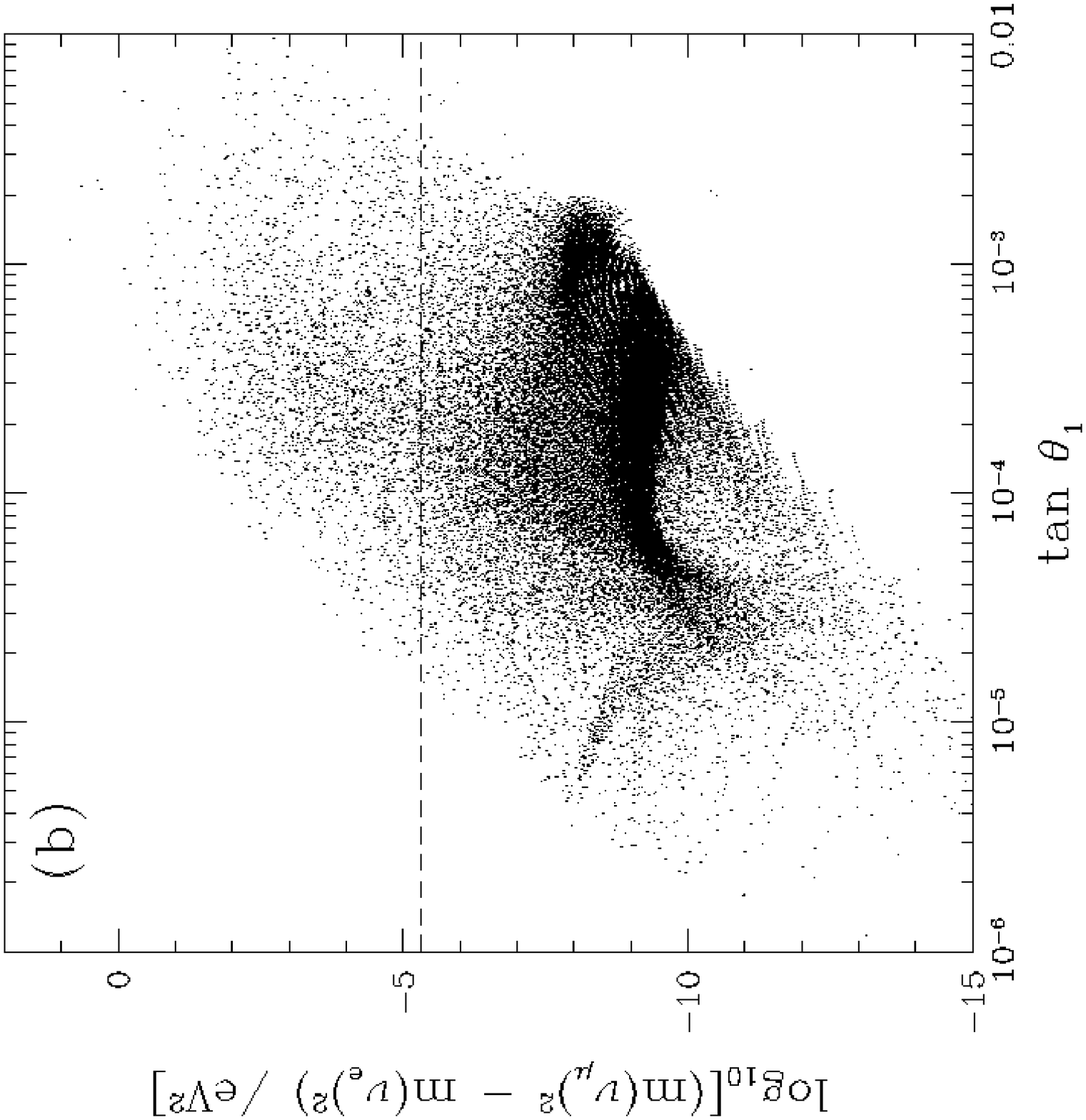}
\caption{
Scatter-plot of $6\times 10^{4}$ models with maximal mixing vs.
(a) log$_{10}(m_0/\gev)$ and
(b) $\tan \theta_1$
for $\sin^2 2\theta_{e\nu_\mu} = 0.008$}
\label{fig6}
\end{figure}

We begin our numerical analysis for the
case of small mixing (\ie\ $\sin^2 2\theta_{e\nu_\mu} = 0.008$).
In fig.~\ref{fig6} we present the prediction of
$m_{\nu_{\mu}}^2- m_{\nu_{e}}^2$
obtained by scanning over the entire SUSY parameter space.
The dashed line indicates the value required to solve
the solar neutrino puzzle. 
The SUSY parameters are chosen as described in the last section.
We present the prediction of
$m_{\nu_{\mu}}^2- m_{\nu_{e}}^2$ as a function of (a) $m_0$ and
(b) $\tan\theta_1$. The value corresponding to a MSW solution of the 
solar neutrino problem is indicated by a dashed line. 
We see that the value of $m_{\nu_{\mu}}^2- m_{\nu_{e}}^2$
is uncorrelated with $m_0$ and grows with $\tan \theta_1$.
By requiring $m_{\nu_{\mu}}^2- m_{\nu_{e}}^2 = 5\times 10^{-6}~\ev^2$
we find that $2\times10^{-5}\lsim \tan\theta_1 \lsim 5\times 10^{-3}$.

\begin{figure}
\vspace*{13pt}
\vspace*{3.2truein}             
\includegraphics{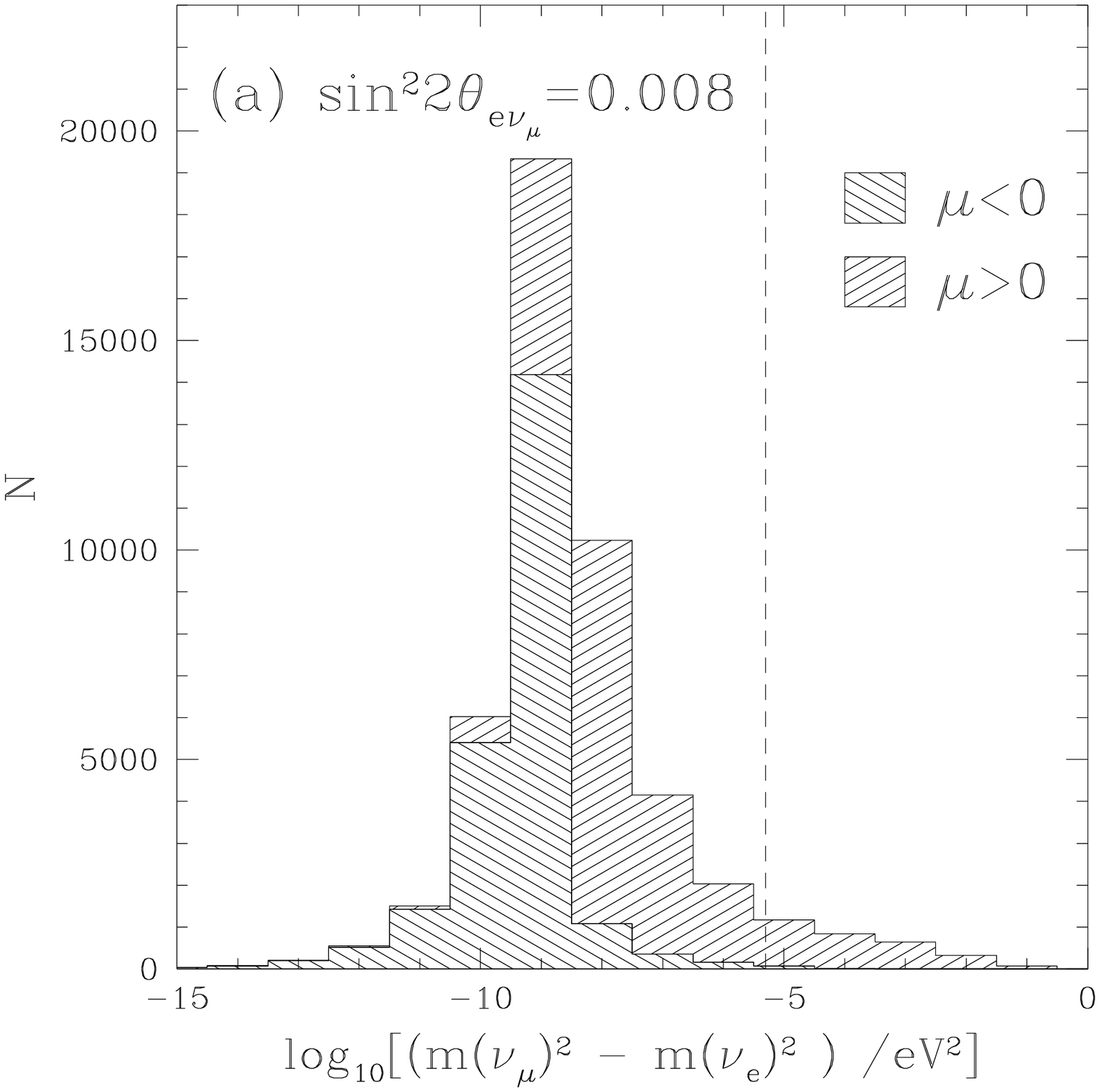}
\includegraphics{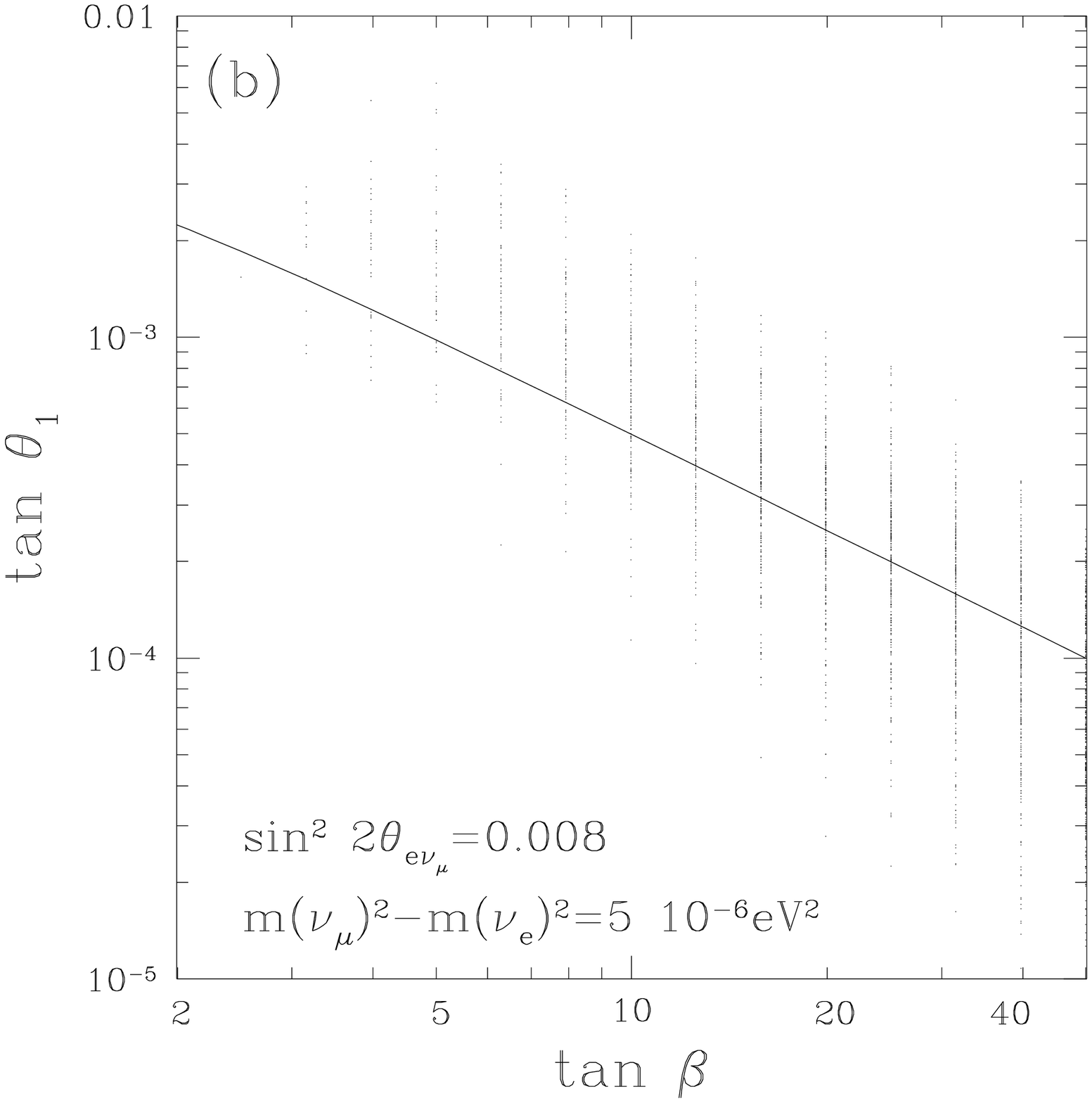}
\caption{
Histogram
of the number of models that yield a particular prediction
for $m_{\nu_{\mu}}^2- m_{\nu_{e}}^2$ and
a scatter plot of all the model 
in the $\tanb$--$\tan\theta_1$ plane. We use the same
models as in fig~\ref{fig6}.
The solid curve in (b) shows the upper limit on the
$R$-parity violating parameter from cosmology
}
\label{fig7}
\end{figure}

In fig.~\ref{fig7} we present a histogram
of the number of models $N$ that yield a particular prediction
for $m_{\nu_{\mu}}^2- m_{\nu_{e}}^2$. 
We have used the same set of models
as in fig.~\ref{fig6}. 
The dashed (dotted) line indicates the value required to solve
the solar neutrino puzzle via the MSW-effect (LWO). 
We see that the predicted value of $m_{\nu_\mu}$
in most models is about one order of magnitude below the
value needed to solve the solar neutrino problem
(the value of  $m_{\nu_e}$
is negligible in all models).
However, the number of models that predict the right value
is still so large that no particular fine-tuning is needed.

In fig.~\ref{fig7}(b) we have singled out all the models with 
$3\times 10^{-6}~\ev^2
 < m_{\nu_{\mu}}^2- m_{\nu_{e}}^2<
 7\times 10^{-6}~\ev^2$.
We find a clear correlation 
between $\tanb$ and $\tan\theta_1$ that allows to predict 
$\tan\theta_1$ to within a factor of three for a give value of
$\tanb$ independent of the SUSY spectrum.
Let us now compare the obtained value for
$\tan\theta_1$ with experimental constraints.
The most solid constraints come from collider experiments. However,
they are very weak\cite{pilaftsis}
and do not constrain the region in parameter
space we are interested in.
Stronger constraints can be obtained by requiring that the
baryon asymmetry not be washed out at the electro-weak phase
transition\cite{cosmology-c}.
It means that at least one of the three lepton has to stay out of
chemical equilibrium, ie. the lepton number violating
Yukawa couplings for the electron have to satisfy
\beqn
y^L_{1 2 1} = {g m_\mu \over \sqrt{2} \mw}{\mu_3\over \mu_0}
= 6\, 10^{-4} {\tan \theta_1 \sin \theta_2 \sin \theta_3 \over \cosb}
\lsim 10^{-7}\,,
\label{c-constraint}
\eeqn
Note that eq.~\ref{c-constraint} only constraints the product
$\tan \theta_1 \sin \theta_2
\propto m_{\nu_\tau}  \sin \theta_2 \propto m_{\nu_\mu}$.
In fig.~\ref{fig7}(b) we see that the small mixing solution to the
solar neutrino problem is compatible with
eq.~\ref{c-constraint} (soid curve)
in particular in the region of large $\tanb$.

\begin{figure}
\vspace*{13pt}
\vspace*{6.5truein}             
\includegraphics{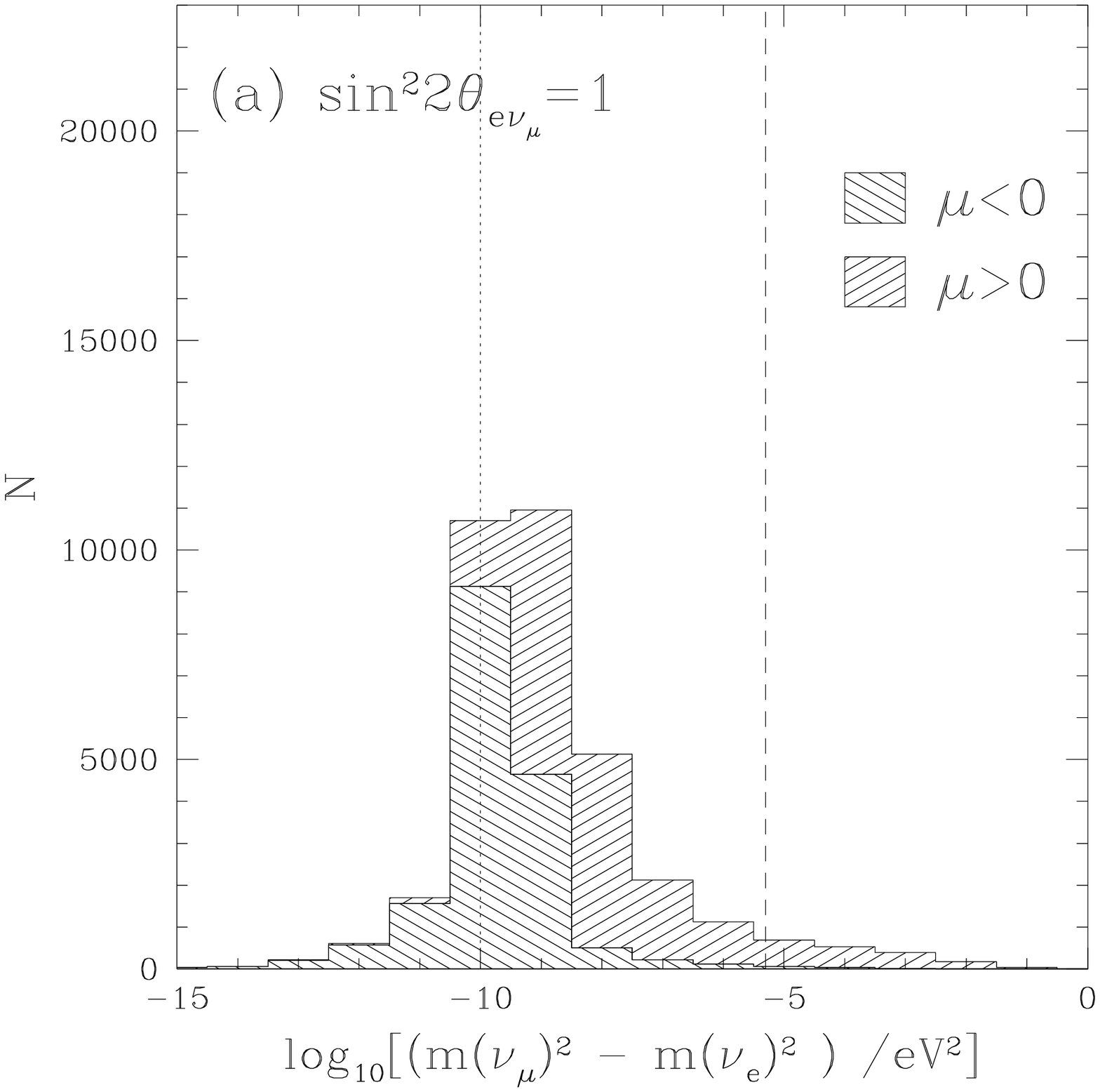}
\includegraphics{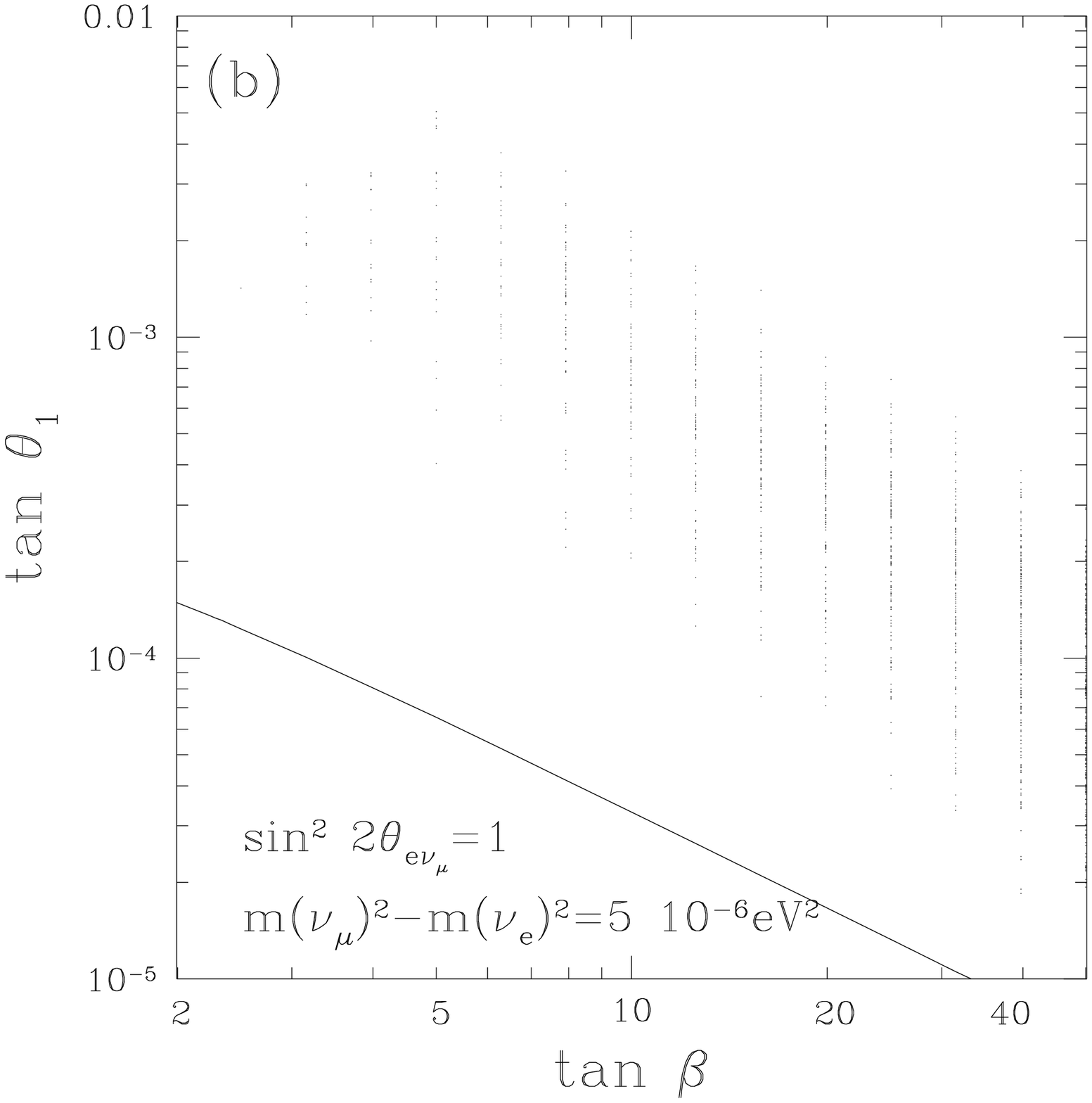}
\includegraphics{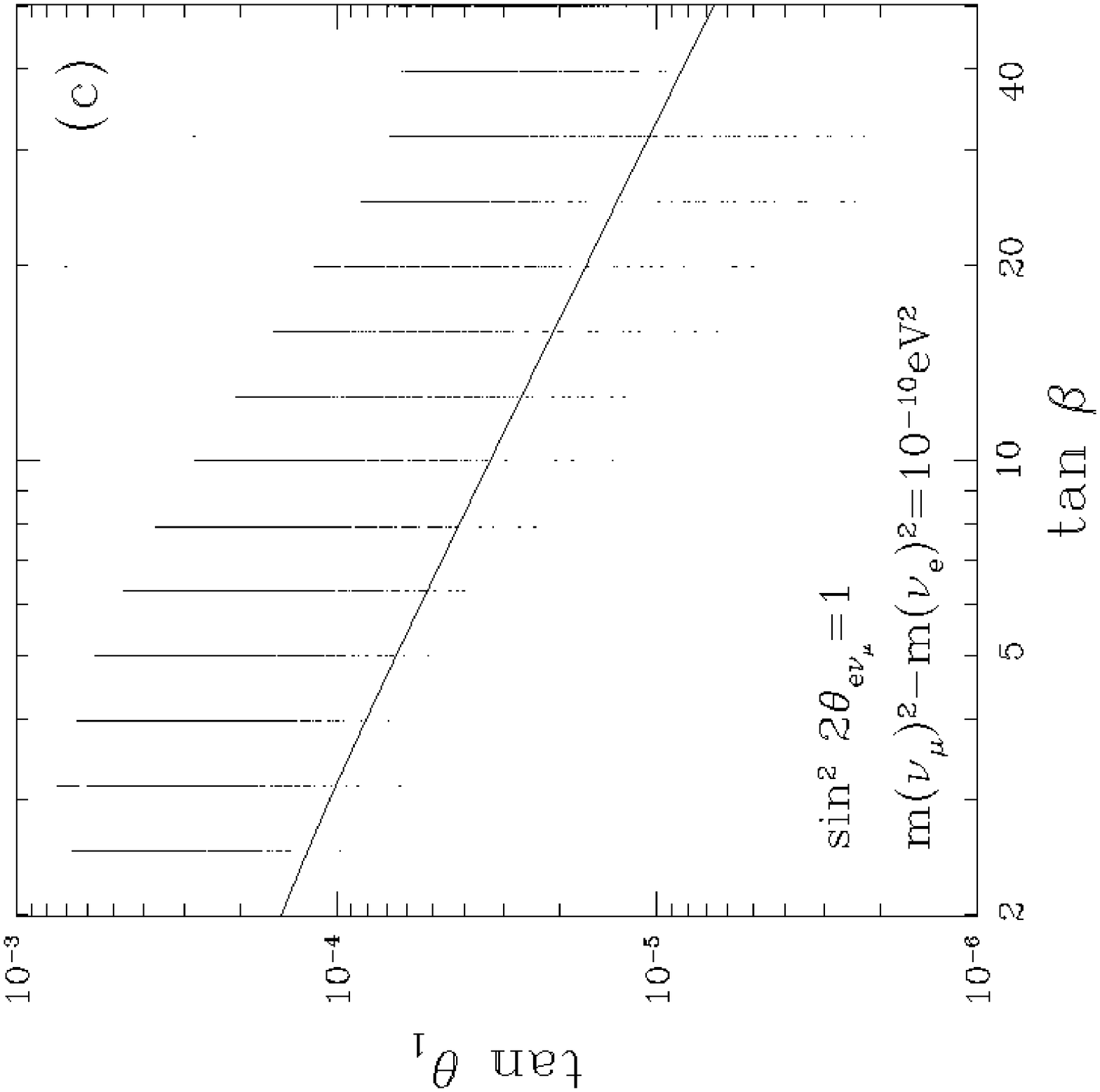}
\caption{
Same as in fig~\ref{fig5} except 
$\sin^2 2\theta_{e\nu_\mu} \simeq 1$
and (b) $m_{\nu_{\mu}}^2- m_{\nu_{e}}^2=5 10^{-5}~\ev^2$
and (c) $m_{\nu_{\mu}}^2- m_{\nu_{e}}^2= 10^{-10}~\ev^2$.
Note that maximum $\nu_e$--$\nu_\mu$ mixing leads
to much stronger constraints on $\tan \theta_1$}
\label{fig8}
\end{figure}

We will now investigate the region of large $\nu_e$--$\nu_\mu$
mixing \ie we adjust $\tan \theta_3$ such that
$\sin^2 2\theta_{e\nu_\mu} \simeq 1$.
This case yields a spectrum of squared mass differences
very similar to the previous case. The result is summarized
in fig.~\ref{fig8}.
We see that the value of $m_{\nu_\mu}^2-m_{\nu_e}^2$
needed for an LWO explanation of the solar neutrino puzzle
lies very close to the maximum.
It is obtained by roughly ten times as many models
as the value needed for an MSW explanation.
Of course it is impossible to draw any rigorous
conclusion from such a scan since we have no objective way
of weighing the probability for a particular set of parameters.
However, in fig.~\ref{fig7}(b) we see that the large mixing MSW solution
is incompatible with cosmological constraints
eq.~\ref{c-constraint} (soid curve)
while for some sets of parameters the large mixing LWO solution
is allowed.

\section{LSND and hot dark matter}

In the last section we answered the question whether
the solar and atmospheric neutrino problems could
be solved simultaneously within the frame-work of our model.
However, there are several other experimental
hints that indicat the existence of neutrino oscillations.
Maybe the most significant result
arises from $\mu$ decay at rest
which requires $m_{\nu_\mu}^2-m_{\nu_e}^2 \gsim 0.3~\ev^2$
and $\sin^2 2\theta_{e\nu_\mu} \simeq 0.004$\cite{lsnd}.
We perform a similar scan over the parameter space
as in fig.~\ref{fig7} and fig.~\ref{fig8}.
Note that here the solar neutrino deficiency
can be explaine by $\nu_e$--$\nu_\tau$
oscillation.
Thus, we fix
$\tan \theta_1$ by tuning
$m_{\nu_\tau}^2-m_{\nu_e}^2 = 5\times 10^{-6}~\ev^2$
$\tan \theta_2$ by tuning $\sin^2 2\theta_{e\nu_\mu} \simeq 0.004$ and
and $\tan \theta_3$ by tuning $\sin^2 2\theta_{e\nu_\tau} \simeq 0.008$.

\begin{figure}
\vspace*{13pt}
\vspace*{3.5truein}             
\includegraphics{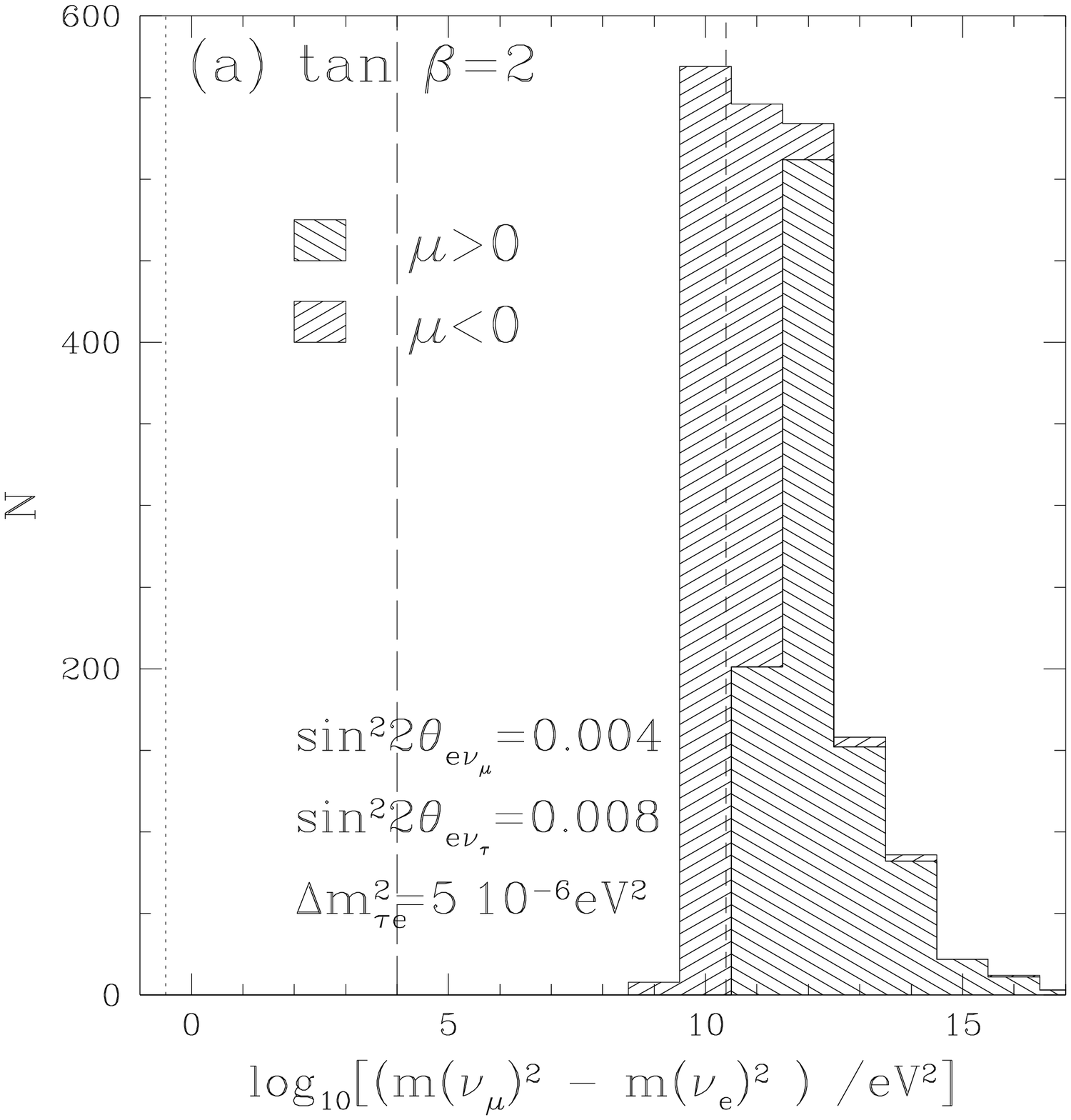}
\includegraphics{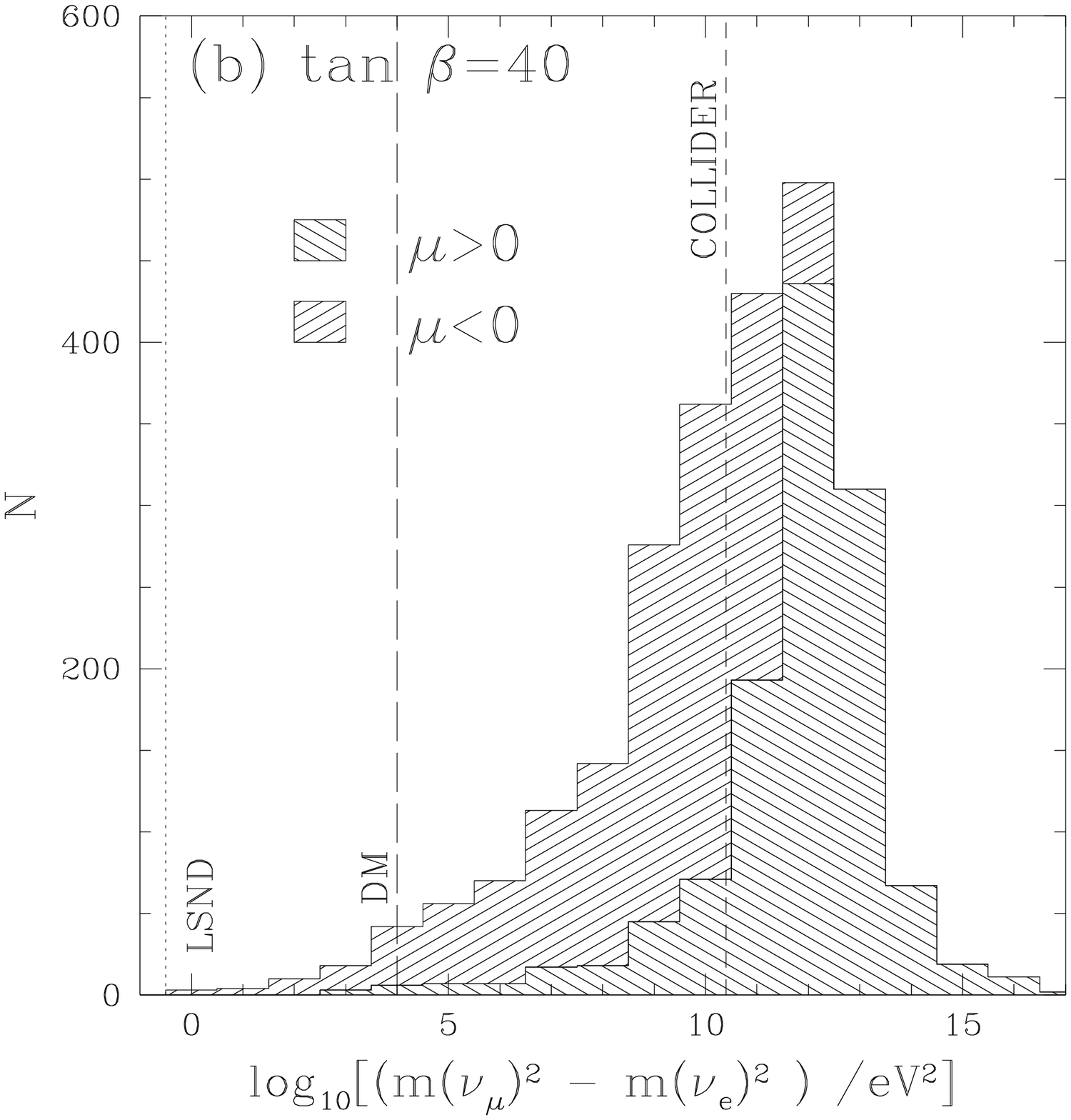}
\caption{
Histogram
of the number of models that yield a particular prediction
for $m_{\nu_{\mu}}^2- m_{\nu_{e}}^2$. We use the same
set of parameters as in fig~\ref{fig5} except
that we only consider $\tanb = 2$ and $\tanb = 40$.
We fix $\tan \theta_2$ by tuning $\sin^2 2\theta_{e\nu_\mu} \simeq 0.004$
and $\tan \theta_3$ by tuning $\sin^2 2\theta_{e\nu_\tau} \simeq 0.008$}
\label{fig9}
\end{figure}

Our result is summarized
in fig.~\ref{fig9}.
The dotted curve shows the lower limit on
$m_{\nu_\mu}^2-m_{\nu_e}^2$ compatible with the LSND result.
The short dashed curve shows the upper limit on
$m_{\nu_\mu}$ from collider experiments\cite{pdg}
while the long dashed  curve shows the upper limit on
the heavies neutrino mass from requiring that the
neutrino relic density does not overclose the universe\cite{k&t}.
We see that for large values of $\tanb$ there are
a few sets of parameters
that can simultaneously explain the solar neutrino deficiency
and the LSND result. However, since in this scenario
$m_{\nu_\mu}$ is quite large in comparison with
$m_{\nu_\tau}$ in the case of the atmospheric neutrino
puzzle it is clear that the $R$-parity violating parameter
is larger than its cosmological upper limit~\cite{cosmology-c}.

One of the disadvantages of models with broken $R$-parity is
the instability of the lightest supersymmetric particle
which cannot serve as a cold dark matter candidate.
One the other hand, it has been suggested that
the existence of hot dark matter (HDM) consisting of
massive neutrinos with a sum of all the masses of about $7$~eV
would be desirable\cite{hdm}.
Clearly, this condition is only compatible with
both a solution to solar and atmospheric neutrino puzzle if
all three neutrinos are almost mass-degenerate.
This is incompatible with the hierarchical spectrum of our model (ie.
$m_{\nu_e}\ll m_{\nu_\mu}, m_{\nu_\tau}$).
However, it is clearly possible to solve the solar neutrino puzzle
via $\nu_e$--$\nu_\mu$ oscillation
while fixing $m_{\nu_\tau} = 7~\ev$.
E.g, take all the models from fig.~\ref{fig7}(b) and
fig.~\ref{fig8}(b)-(c) and replace
$\tan \theta_1 \rightarrow (7~\ev/0.1~\ev)^{1/2}\tan \theta_1$ and
$\tan \theta_2 \rightarrow (7~\ev/0.1~\ev)^{-1/2}\tan \theta_2$.
This rescaling only changes the value of $m_{\nu_\tau}$
while the values of $m_{\nu_\mu}$, $m_{\nu_e}$
and the $R$-parity violating
coupling constrained by eq.~\ref{c-constraint} remain the same.

\chapter{Conclusions}

We have investigated the spectrum of neutrino masses and mixing angles
in the MSSM with broken $R$-parity.
We have focused on the model where 
$R$-parity breaking arises explicitly from
dimension 2 terms in the superpotential. This model is
characterized by only three additional parameters
and can be embedded
in SUSY GUT model without any constraints from
the proton life-time.
In this model, the $R$-parity breaking is characterized by 
$\tan\theta_1$, the $\mu$--$\tau$ mixing by $\tan \theta_2$,
and the $e$--$\mu$ mixing by $\tan \theta_3$.
By adjusting these parameters we can 
solve the solar and atmospheric neutrino puzzles
without fine-tuning.
In particular, we have shown that a large hierarchy
$m_{\nu_\tau} \gg m_{\nu_\mu} \gg m_{\nu_e}$
is quite natural even in the case of maximum mixing
and may favor a LWO explanation of the solar neutrino puzzle.
To obtain the correct neutrino masses we need
$\tan\theta_1\simeq 10^{-x}$  ($x = 2\sim 5$;
this ratio may be explained as the ratio of $\mgut/\mpl$) but no
additional particles and no new intermediate scale.
Requiring that the baryon asymmetry not be washed out at the
electro-weak phase transition rules out the large mixing MSW solution,
while some regions of the
LWO solution and small mixing MSW solution are allowed.

Finally, we demonstrate that there are regions in parameter space
where the solution to the solar neutrino puzzle
is compatible with either the LSND result or
the existence of a $7~\ev$ HDM neutrino.

\underline{Acknowledgement:} I would like to thank
D. Pierce for comparison of the numerical results,
S. Davidson, G. Raffelt and N. Polonsky
for many useful and pleasant conversations
and the ITP in Santa Barbara where this work was completed
for their kind hospitality.

This work was supported in parts by the 
National Science Foundation Grant
No. PHY94-07194.

\begin{appendix}

\chapter{The RGEs for the $R$-parity violating terms}

Here we present the RGEs for the $R$ symmetry breaking
terms. They can easily be derived from ref.~\cite{falck}
In our model without 
$R$-parity violating Yukawa
couplings and with diagonal lepton Yukawa couplings
the RGEs for $\mu_I$ reduce
to\footnote{After completion of this paper,
a complete set of RGEs was presented
in ref.~\cite{dreinerrge}}
\beqn
32 \pi^2{\d \mu_0\over \d t} &=&  
\mu_0 \left( \tr y^{L2} +  N_c\tr y^{D2} +  
N_c\tr y^{U2} -3 g^2 - g^{\prime2} 
\right)\,,\nonumber\\
32 \pi^2{\d \mu_i\over \d t} &=&
\mu_i \left( y^{L}_{(i i)}y^L_{(i i)} 
 +  N_c\tr y^{U2} -3 g^2 - g^{\prime2} 
\right)\,,
\label{rgemup}
\eeqn
where we do not sum over indices in brackets.
Furthermore, for our purposes it is more convenient to write
$B_I^\mu = B_{I J} \mu_J$ with
\beqn
16 \pi^2{\d B_{0 0}\over \d t} &=&
\tr A^L y^L + N_c \tr A^D y^D + N_c \tr A^U y^U
+ 3 M g^2 + M^\prime g^{\prime 2}\,,\nonumber\\
16 \pi^2{\d B_{(i i)}\over \d t} &=&
A^L_{(i i)} y^L_{(i i)} + N_c \tr A^U y^U
+ 3 M g^2 + M^\prime g^{\prime 2}\,,
\label{rgemip}
\eeqn

\chapter{The mass at tree-level}

In our model the electro-weak symmetry is broken via a
the vacuum expectation value (VEV) of the neutral CP-even Higgs boson
fields and the sneutrino fields, $\bar v \equiv \vev{H}$
and $v_I \equiv \vev{L_I}$. We find it convenient to present
all the masses and interactions in the basis
where the interactions are $R$-parity conserving
and $R$-parity breaking is only parameterized by
the mass parameters $\mu_i$, $B^\mu_i$, and after electroweak
symmetry breaking also by $v_i$.
Furthermore, it is convenient to define the unit vector
$u_I \equiv v_I/v$.

\begin{table}[t]
$$
\begin{array}{|c||c|c|c|c|c|c|}
\hline
\Phi    &     Q     &   U      &     D    &  L & E & H
\\ \hline \hline
Y_\Phi & {1\over 3} & {4\over 3} & -{2\over 3} & -1 & -2 & 1
\\ \hline
\end{array}
$$
\caption{
The hypercharges of the Higgs fields and the matter fields.
Note that $Y_{\Phi^c} = -Y_{\Phi}$ for $\Phi = U, D, E$
}
\label{ytable}
\end{table}

In our notation we use calligraphic letters to denote
the mass matrices and interaction matrices in the electro-weak basis
and roman letters for
the mass matrices and interaction matrices in the basis of
the mass eigenstates. The hypercharges 
are listed in table~\ref{ytable}.

In models with broken $R$-parity the neutrinos mix with the 
neutralinos. We can write the potential as
\beqn
V = \half \left(\matrix{\psi^0}\right) \calm_{\chi^0}
  \left(\matrix{\psi^0}\right)^T
+ {\rm H.c.}\,,\label{blabla}
\eeqn
where $\psi^0_n = \left(-i \lambda^\prime, -i \lambda^3,
\psi_H^2, \psi_{L_I}^1\right)$\footnote{%
Here, the superscripts are the SU(2)$_L$ indices.
Note that we have switched the third and fourth 
component as opposed to other
authors\cite{haberkane} for convenience.}
and the $7\times7$ neutrino/neutralino mass matrix
\beqn
 \calm_{\chi^0} = \left(
\matrix{M^\prime & 0     & \mz \sw\sb     &-\mz \sw\cb u_I \cr
         {}  &M    &-\mz \cw\sb     & \mz \cw\cb u_I \cr
     {}    & {}    & 0 & -\mu_I     \cr
    {} & {\rm symmetric}& {}&  0 \cr
}\right)
\,.\label{defyr}
\eeqn
where we have defined $\cw\equiv \cos\theta_{\rm w} \equiv \mw/\mz$
and $\tan\beta \equiv \bar v/ v$, $\sb \equiv \sin \beta$, etc. 
We now define the unitary matrix $Z$ such that the matrix
$m_{\chi^0} = Z^*\calm_{\chi^0}Z^{-1}$ is diagonal
and the value of the diagonal elements is
ordered by magnitude.

The mass matrix in eq.~\ref{defyr} still has two massless eigenstates
corresponding to two neutrinos. However, one neutrino acquires a mass 
\beqn
m_{\nu_3} \simeq { \cb^2 \left(M \sw^2 + M^\prime \cw^2\right) \mz^2 
\mu^2 \sin^2 (\theta_1-\theta_1^\prime)
\over 
 M^\prime M \mu_0^2 
- 2\sb\cb \left(M \sw^2 + M^\prime \cw^2\right) \mz^2 \mu_0}
\,,
\label{mtau}
\eeqn
at tree-level.
Furthermore, the charged leptons mix with the charginos.
In complete analogy to the MSSM we can write
\beqn
\psi^+ &=& \left(-i\lambda^+, \psi_{H}^1, \psi_{E^c_i}\right)\nonumber\\
\psi^- &=& \left(-i\lambda^-, \psi_{L_I}^2\right)
\eeqn
The potential is given by
\beqn
V = \psi^-\calm_{\chi^\pm} \psi^+ + \hc
\eeqn
The $5\times 5$ mass matrix is given by
\beqn
{\cal M}_{\chi^\pm} = \left(
\matrix{M          & \sqrt{2}\mw\sinb      & 0 \cr
      \sqrt{2}\mw\cosb u_0  & \mu_0    &-y^L_{i j} v_i\cr
      \sqrt{2}\mw\cosb u_i  & \mu_i    & y^L_{i j} v_0\cr
}\right)\,,
\label{defxr}
\eeqn
We now define the unitary matrices $U^R$, $U^L$ such that
$m_{\chi^\pm} \equiv U^{R*} \calm_{\chi^\pm} (U^L)^{-1}$
is diagonal (in the MSSM
$U^R = U$ and $U^L = V$).
Then we can write the chargino mass eigenstates as
four component spinors
\beqn
\chi^+ = \left(\matrix{U^L \psi^+
                   \cr U^R \psibar^-}\right)\,,
\eeqn

Let us consider the Higgs sector.
In the case of explicitly broken $R$-parity (i.e. $\mu_i \neq 0$)
we find that the Higgs multiplets mix with the lepton multiplets.
In particular, the sneutrinos acquire a non-zero VEV. The 
system of five equation corresponding to the minimum conditions of these
fields can be solved most easily numerically by iteration.
The tree-level mass matrix of the neutral, 
CP-even Higgs/slepton mass matrix
is given by
\beqn
\calm_{H^0} = \left(\matrix{\mu_I^2 + m_H^2 + \half \mz^2(3\sb^2-\cb^2)
    &  -B_J^\mu - \mz^2\sb\cb u_J
   \cr -B_I^\mu - \mz^2\sb\cb u_I
    & \mu_I \mu_J + m_{L_{I J}}^2
  + \half \mz^2[\cb^2 (2 u_I u_J + \delta_{I J})-\sb^2\delta_{I J}]
   }\right)\,,
\label{massh0}
\eeqn
and tree-level mass matrix of the neutral, 
CP-odd Higgs/slepton mass matrix
is given by
\beqn
\calm_{A^0} &=& \left(\matrix{\mu_I^2 + m_H^2 + \half \mz^2(\sb^2-\cb^2)
    &  -B_J^\mu
   \cr -B_I^\mu
    & \mu_I \mu_J + m_{L_{I J}}^2
  + \half \mz^2(\cb^2 - \sb^2) \delta_{I J}
   }\right)\nonumber\\
&&+ \xiz\mz^2 \left(\matrix{\sin^2\beta& \sinb\cosb u_J\cr 
   \sinb\cosb u_I & \cos^2\beta u_I u_J}\right)
\,,
\label{massa0}
\eeqn
Finally, the two charged Higgs bosons mix with the
charged sleptons. The mass matrix in the electro-weak
eigenbasis $H_a^+ = (H^+, \widetilde E^{L+}_I, \widetilde E^{R+}_i)$
with $a = 1,...,8$ becomes
 \beqn
{\cal M}_{H^\pm}^2 = 
\left(\matrix{{\cal M}_{L L}^2&{\cal M}_{L R}^2\cr
{\cal M}_{R L}^2&{\cal M}_{R R}^2}\right)\,,
\eeqn
where $\widetilde E^{L+} = (\widetilde L^-)^*$,
$\widetilde E^{R+} =  \widetilde E^c$
where the submatrices are given by
\beqn
{\cal M}_{L L}^2 &=& 
\left(\matrix{M_{H}^2 + \mu^2 & -B^\mu_J \cr
 -B^\mu_I & M_{L_{I J}}^2 + \mu_I \mu_J + M_{I J}^{y 2}}\right)
-\half \mz^2 \sw^2 \cos 2\beta
\left(\matrix{Y_H&0\cr 0&Y_L\delta_{I J}}\right)\nonumber\\
&&+\half \mw^2 \left(\matrix{1&-\sin 2\beta u_J\cr 
 - \sin 2\beta u_I &2 \cos^2\beta u_I u_J 
- \cos 2\beta\delta_{I J}}\right)\nonumber\\
&&+ \xiw\mw^2 \left(\matrix{\sin^2\beta& \sinb\cosb u_J\cr 
   \sinb\cosb u_I & \cos^2\beta u_I u_J}\right)
\eeqn
with $M_{I J}^{y 2}$ given by
\beqn
M_{0 0}^{y 2} &=& M_{0 j}^{y 2} = M_{i 0}^{y 2} = 0\,,\nonumber\\
M_{i j}^{y 2} &=&
 y^L_{i k}y^L_{j k} v_0^2
+y^L_{i k}y^L_{j l} v_k v_l\,,
\eeqn
and
\beqn
{\cal M}_{L R}^2 = 
\left(\matrix{ y_{i j}^L (v_0 \mu_i - v_i\mu_0)
\cr A_{i j}^L v_i - \mu_i y_{i j}^L \barv
\cr-A_{i j}^L v_0 + \mu_0 y_{i j}^L \barv}\right)
\eeqn
\beqn
{\cal M}_{R R}^2 = M_{E_{i j}}^2
-\half Y_{E^c} \mz^2 \sw^2 \cos 2\beta
+y^L_{k i}y^L_{k j} v_0^2
+y^L_{k i}y^L_{l j} v_k v_l
\eeqn
Note that the gauge dependent pieces arise from the gauge fixing terms
in the $R_\xi$ gauge.
They give masses to the goldstone-bosons
$m_{A^0_1}^2   = \xiz \mz^2$ and
$m_{H^\pm_1}^2 = \xiw \mw^2$, respectively,
Inside our one-loop calculation of Appendix~D these
unphysical fields are treaded just like the physical
fields.

\chapter{Vertices}

In the following appendix we summarize a set of vertices
in a notation that is convenient for SUSY theories with broken $R$
parity[\cite{pilaftsis},\cite{yossi}].
Note, that due to the minus signs in eq.~2.6
the fields $\psi_x^-$ defined in eq.~B.4
are minus the fermionic 
partners of $H_x^-$ ($x = 2, 3, 4, 5$)
defined above eq.~B.10.

The Feynman rules are given by
$i \gamma_\mu (V^L L +V^R R)$,
$-i(V^L L +V^R R)$,
$-(V^L L +V^R R)$,
for couplings of a fermion pair to a gauge boson, a
complex or CP-even scalar, and a CP-odd scalar,
respectively (here, $V^L$ and $V^R$ is the generic notation for
all the vertices listed below. Note that we have included the
gauge coupling constant in the definitions of $V^{L/R}$).

The $W^- \chi^0_n\chi^+_x$ vertices are given by ($n = 1,..,7$ and
$x = 1,..,5$)
\beqn
\calo^{L}_{n x} 
&=& g \left(\delta_{n 2}\delta_{x 1}
      -{1\over \sqrt{2}}\delta_{n 3}\delta_{x 2}\right)\,,\nonumber\\
\calo^{R}_{n x} 
&=& g \left(\delta_{n 2}\delta_{x 1}
      +{1\over \sqrt{2}}\sum_{\alpha = 2}^5
      \delta_{n \alpha+2}\delta_{x \alpha}\right)\,.
\eeqn
The $Z \chi^-_x\chi^+_y$ vertices are given by
\beqn
\calo^{\prime L} 
&=& {g\over 2\cw}\diag \left(2\sw^2 -2, 2\sw^2 -1 , 2 \sw^2\delta_{(i i)}
\right)\,,\nonumber\\
\calo^{\prime R}
&=& {g\over 2\cw}\diag \left(2\sw^2 -2, 
(2\sw^2 -1)\delta_{(I I)}\right)\,,
\eeqn
no summation over indices in parenthesis.
The $Z \chi^0_m\chi^0_n$ vertices are given by
\beqn
\calo^{\prime\prime L} 
&=&  {g\over 2\cw} \diag \left(0,0,1, - 
\delta_{(I I)}\right)\,,\nonumber\\
\calo^{\prime\prime R} &=& -\calo^{\prime\prime L} 
\eeqn
The $H^0_x \chi^-_y\chi^+_z$ vertices are given by
\beqn
\calq^{L}_{x y z}
= {1\over \sqrt{2}}\times \cases{
 g \delta_{1 y} \delta_{z 2}
&for  $x = 1$\cr
 g \delta_{2 y} \delta_{z 1}
+y^E_{y-2 z-2}
&for  $x = 2$\cr
 g \delta_{x y} \delta_{z 1}
 -y^E_{x-2 z-2} \delta_{y 2}
&for  $x = 3,4,5,$}
\eeqn
and $\calq^{R}_{x y z} = \calq^{L}_{x z y}$.
The $A^0_x \chi^-_y\chi^+_z$ vertices are given by
\beqn
\cals^{L}_{x y z}
= \cases{
\calq^{L}_{x y z}
&for  $x = 1$\cr
-\calq^{L}_{x y z}
&for  $x = 2,3,4,5,$}
\eeqn
and $\cals^{R}_{x y z} = -\cals^{L}_{x z y}$.
The $H^0_x \chi^0_m\chi^0_n$ vertices are given by
\beqn
\calq^{\prime\prime P}_{x m n}
= \cases{
-{g\over 2}\left[\delta_{m x+2}
\left(\delta_{n 2}-\tanw\delta_{n 1}\right)
           + \delta_{n x+2}\left(\delta_{m 2}-\tanw\delta_{m 1}\right)
\right]  &for  $x = 1$\cr
{g\over 2}\left[\delta_{m x+2}
\left(\delta_{n 2}-\tanw\delta_{n 1}\right)
           + \delta_{n x+2}\left(\delta_{m 2}-\tanw\delta_{m 1}\right)
\right]  &for  $x = 2,3,4,5$}
\eeqn
where $P = L,R$.
The $A^0_x \chi^0_m\chi^0_n$ vertices are given by
\beqn
\cals^{\prime\prime L}_{x m n} =
-\cals^{\prime\prime R}_{x m n} =
{g\over 2}\left[\delta_{m x+2}
\left(\delta_{n 2}-\tanw\delta_{n 1}\right)
           + \delta_{n x+2}\left(\delta_{m 2}-\tanw\delta_{m 1}\right)
\right]\,,
\eeqn
The $H^-_a \chi^0_n\chi^+_x$ vertices are given by
\beqn
\calq^{\prime L}_{a n x} 
= \cases{
     Y_H {g^\prime\over \sqrt{2}} \delta_{n 1}\delta_{x 2}
    +{g       \over \sqrt{2}} \delta_{n 2}\delta_{x 2}
    + g \delta_{n 3}\delta_{x 1}  &for  $a = 1$\cr
               y_{n-4 x-2}^L & for $a = 2$\cr
              -y_{a-2 x-2}^L\delta_{n 4} & for $a = 3,4,5$\cr
    Y_E {g^\prime\over \sqrt{2}} \delta_{n 1}\delta_{x a-3} & for $a = 6,7,8$}
\eeqn
\beqn
\calq^{\prime R}_{a n x} &=& \cases{0 & for $a = 1$\cr
   -Y_L {g^\prime\over \sqrt{2}} \delta_{n 1}\delta_{x a}
   +{g       \over \sqrt{2}} \delta_{n 2}\delta_{x a}
   - g \delta_{n a+2}\delta_{x 1} & for $a = 2,..,5$\cr
           -y_{n-4 a-5}^L\delta_{x 2}
           +y_{x-2 a-5}^L\delta_{n 4}  & for $a = 6,7,8$
}
\eeqn
The $\tilde u_\alpha \bar u_b\chi_n^0$ vertex is
\beqn
\calv_{\alpha b n}^{u L} &=& y^U_{\alpha b} \delta_{n 3}
 -{1\over \sqrt{2}}Y_U g^\prime 
\delta_{\alpha-3 b}\delta_{n 1}\nonumber\\
\calv_{\alpha b n}^{u R} &=& y^U_{b \alpha-3} \delta_{n 3}
 +{1\over \sqrt{2}} g \delta_{\alpha b}\delta_{n 2}
 +{1\over \sqrt{2}}Y_Q g^\prime \delta_{\alpha b}\delta_{n 1}
\eeqn
The $\tilde d_\alpha \bar d_b \chi_n^0$ vertex is
\beqn
\calv_{\alpha b n}^{d L} &=& y^D_{\alpha b} \delta_{n 4}
 -{1\over \sqrt{2}}Y_D g^\prime \delta_{\alpha-3 b}\delta_{n 1}\nonumber\\
\calv_{\alpha b n}^{d R} &=& y^D_{b \alpha-3} \delta_{n 4}
 -{1\over \sqrt{2}} g \delta_{\alpha b}\delta_{n 2}
 +{1\over \sqrt{2}}Y_Q g^\prime \delta_{\alpha b}\delta_{n 1}
\eeqn
The $\tilde u_\alpha \bar d_b\chi_x^-$ vertex is
\beqn
\calv_{\alpha b x}^{\tilde u L} C&=& -y^D_{\alpha b} \delta_{x 2} C
\nonumber\\
\calv_{\alpha b x}^{\tilde u R} C &=&
 \left( g \delta_{\alpha b}\delta_{x 1}
 - y^U_{b \alpha-3} \delta_{x 2} \right) C
\eeqn
The $\tilde d_\alpha \bar u_b \chi_x^+$ vertex is
\beqn
\calv_{\alpha b x}^{\tilde d L}  &=&
 - y^U_{\alpha b} \delta_{x 2}
\nonumber\\
\calv_{\alpha b x}^{\tilde d R} &=&
 g \delta_{\alpha b}\delta_{x 1}
 - y^D_{b \alpha-3} \delta_{x 2}
\eeqn
Definition and more details on the charge conjugation operator
$C$ can be found in ref.~\cite{haberkane}.

The vertices in the mass eigenbasis are given by
\beqn
O_{n x}^P &=& Z_{n \tilde n} U^P_{x \tilde x}
\calo_{\tilde n \tilde x}^P\nonumber\\
O_{x y}^{\prime P}
&=& U^P_{x \tilde x} U^P_{y \tilde y}
\calo_{\tilde x \tilde y}^{\prime P}\nonumber\\
O_{m n}^{\prime\prime P} &=& Z_{m \tilde m} 
Z_{n \tilde n}\calo_{\tilde m \tilde n}^{\prime\prime P}\nonumber\\
S_{x y z}^L &=&U^{A^0}_{x \tilde x}
U^R_{y \tilde y} U^L_{z \tilde z}
\cals_{\tilde x \tilde y\tilde z}^L\nonumber\\
S_{x y z}^R &=&U^{A^0}_{x \tilde x}
U^L_{y \tilde y} U^R_{z \tilde z}
\cals_{\tilde x \tilde y\tilde z}^R\nonumber\\
Q_{x y z}^L &=&U^{H^0}_{x \tilde x}
U^R_{y \tilde y} U^L_{z \tilde z}
\calq_{\tilde x \tilde y\tilde z}^L\nonumber\\
Q_{x y z}^R &=&U^{H^0}_{x \tilde x}
U^L_{y \tilde y} U^R_{z \tilde z}
\calq_{\tilde x \tilde y\tilde z}^R\nonumber\\
Q_{\alpha n x}^{\prime P}&=&
U^{H^\pm}_{\alpha \tilde\alpha} Z_{n \tilde n} U^P_{x \tilde x}
\calq_{\tilde\alpha \tilde n \tilde x}^{\prime P}\nonumber\\
Q_{x m n}^{\prime\prime P} &=& U^{H^0}_{x \tilde x}
Z_{m \tilde m} 
Z_{n \tilde n}
\calq_{\tilde x\tilde m \tilde n}^{\prime\prime P}\nonumber\\
S_{x m n}^{\prime\prime P} &=& U^{A^0}_{x \tilde x}
Z_{m \tilde m} 
Z_{n \tilde n}
\cals_{\tilde x\tilde m \tilde n}^{\prime\prime P}\nonumber\\
V_{\alpha a x}^{\tilde d P} &=& 
U^{\tilde d}_{\alpha \tilde \alpha} U_{x \tilde x}^{P} 
\calv_{\tilde \alpha a \tilde x}^{\tilde d P}\nonumber\\
V_{\alpha a x}^{\tilde u L/R} &=& 
U^{\tilde u}_{\alpha \tilde \alpha} U_{x \tilde x}^{R/L} 
\calv_{\tilde \alpha a \tilde x}^{\tilde u L/R}\nonumber\\
V_{\alpha a m}^{q P} &=& 
U^{\tilde q}_{\alpha \tilde \alpha} Z_{m \tilde m} 
\calv_{\tilde \alpha a \tilde m}^{q P}
\qquad (q = u, d)\,,
\label{rotations}\eeqn
for $P = L,R$ (here $U^L \equiv V$, and $U^R \equiv U$).
Furthermore, we have assumed that the Yukawa matrices $y^{U/D/L}$
are diagonal. The unitary matrices $U^{\tilde q}$
diagonalizing the squark mass matrices are defined analogous to
eq.~2.15.

\chapter{The One-Loop Corrected Neutrino/Neutralino\\ Mass Matrix}

In this appendix we present the
complete one-loop radiative corrections to
the full neutrino/neutralino mass matrix.
Note that for the radiative generation of the neutrino masses
only the diagrams involving Higgs fields are relevant.
We have regularized the divergences by
dimensional reduction\cite{dimred}.
The calculation has been done in the 
t'Hooft--Feynman gauge ($\xi = 1$).
Note that the self-energy diagrams and as a consequence also
the running $\drbar$ masses as defined in eq.~\ref{drmasses}
are gauge-dependent.
There are seven types of diagrams contributing to the self-energy
of the Neutrino/Neutralino:
charged Higgs/chargino,
CP-even Higgs/neutralino,
CP-odd Higgs/neutralino,
up quark/squark,
down quark/squark,
$W^\pm$/chargino,
$Z$/neutralino. Their results are
\beqn
\Sigma^{V}_{n m}  &=& \half \left(\Sigma^R_{n m} + \Sigma^L_{n m} \right)
\nonumber\\
                    &=&-{1\over 16 \pi^2}\left[
 \sum_{a,x}\left(Q^{\prime L}_{a n x} Q^{\prime L}_{a m x}
                +Q^{\prime R}_{a n x} Q^{\prime R}_{a m x}\right) 
                   B_1(p^2,m_{\chi^\pm_x}^2,m_{H^\pm_a}^2)
            \right.\nonumber\\
&&+\half\sum_{x,k}\left(Q^{\prime\prime L}_{x n k}
 Q^{\prime\prime L}_{x m k}
 +Q^{\prime\prime R}_{x n k} Q^{\prime\prime R}_{x m k}\right)
                   B_1(p^2,m_{\chi^0_k}^2,m_{H^0_x}^2)\nonumber\\
&&+\half\sum_{x,k}\left(S^{\prime\prime L}_{x n k} 
S^{\prime\prime L}_{x m k}
 +S^{\prime\prime R}_{x n k} S^{\prime\prime R}_{x m k}\right)
                   B_1(p^2,m_{\chi^0_k}^2,m_{A^0_x}^2)\nonumber\\
&&+3 \sum_{\alpha, b}\left(V^{u L}_{\alpha b n} V^{u L}_{\alpha b m}
                +V^{u R}_{\alpha b n} V^{u R}_{\alpha b m}\right) 
                   B_1(p^2,m_{u_b}^2,m_{\tilde u_\alpha}^2)\nonumber\\
&&+3 \sum_{\alpha, b}\left(V^{d L}_{\alpha b n} V^{d L}_{\alpha b m}
                +V^{d R}_{\alpha b n} V^{d R}_{\alpha b m}\right) 
                   B_1(p^2,m_{d_b}^2,m_{\tilde d_\alpha}^2)\nonumber\\
&&+2 \sum_{x}\left(O^{L}_{n x} O^{L}_{m x}
                +O^{R}_{n x} O^{R}_{m x}\right) 
                   B_1(p^2,m_{\chi^\pm_x}^2,\mw^2)\nonumber\\
&&\left.
  + \sum_{k}\left(O^{\prime\prime L}_{n k} O^{\prime\prime L}_{m k}
                +O^{\prime\prime R}_{n k}
 O^{\prime\prime R}_{m k}\right) 
                   B_1(p^2,m_{\chi^0_k}^2,\mz^2)
                    \right]\,,\label{defs}\\
\Pi^{V}_{n m}     &=& \half \left(\Pi_{n m}^R   + \Pi^L_{n m}    \right)
\nonumber\\
                    &=&-{1\over 16 \pi^2}\left[
 \sum_{a,x}m_{\chi^\pm_x}\left(Q^{\prime L}_{a n x} Q^{\prime R}_{a m x}
                +Q^{\prime R}_{a n x} Q^{\prime L}_{a m x}\right)
                   B_0(p^2,m_{\chi^\pm_x}^2,m_{H^\pm_a}^2)
                          \right.\nonumber\\
&&+\half\sum_{x,k}m_{\chi^0_k}
    \left(Q^{\prime\prime L}_{x n k} Q^{\prime\prime R}_{x m k}
         +Q^{\prime\prime R}_{x n k} Q^{\prime\prime L}_{x m k}\right) 
                   B_0(p^2,m_{\chi^0_k}^2,m_{H^0_x}^2)\nonumber\\
&&+\half\sum_{x,k}m_{\chi^0_k}
\left(S^{\prime\prime L}_{x n k} S^{\prime\prime R}_{x m k}
     +S^{\prime\prime R}_{x n k} S^{\prime\prime L}_{x m k}\right)
                   B_0(p^2,m_{\chi^0_k}^2,m_{A^0_x}^2)\nonumber\\
&&+3 \sum_{\alpha, b}m_{u_b}
 \left(V^{u L}_{\alpha b n} V^{u R}_{\alpha b m}
      +V^{u R}_{\alpha b n} V^{u L}_{\alpha b m}\right) 
                   B_0(p^2,m_{u_b}^2,m_{\tilde u_\alpha}^2)\nonumber\\
&&+3 \sum_{\alpha, b}m_{d_b}
  \left(V^{d L}_{\alpha b n} V^{d R}_{\alpha b m}
       +V^{d R}_{\alpha b n} V^{d L}_{\alpha b m}\right) 
                   B_0(p^2,m_{d_b}^2,m_{\tilde d_\alpha}^2)\nonumber\\
&&-4 \sum_{x}m_{\chi^\pm_x}
   \left(O^{L}_{n x} O^{R}_{m x}
        +O^{R}_{n x} O^{L}_{m x}\right) 
                   B_0(p^2,m_{\chi^\pm_x}^2,\mw^2)\nonumber\\
&&\left.
  -2 \sum_{k}m_{\chi^0_k}
 \left(O^{\prime\prime L}_{n k} O^{\prime\prime R}_{m k}
      +O^{\prime\prime R}_{n k} O^{\prime\prime L}_{m k}\right) 
                   B_0(p^2,m_{\chi^0_k}^2,\mz^2)\right]\,.
\label{defp}
\eeqn
The $B_0$ and $B_1$ functions are defined by
\beqn
[B_0(q^2,m_1^2,m_2^2),q_\mu B_1(q^2,m_1^2,m_2^2)]
=-16\, i\pi^2 \int{\d^n k\over (2\pi)^n}{[1,k_\mu]\over D}\,,
\eeqn
where $D\equiv (k^2-m_1^2+i\delta)[(k+q)^2-m_2^2+i\delta]$.

\end{appendix}


\end{document}